\documentclass[a4paper, 11pt, oneside]{article} % A4 paper size, default 11pt font size and oneside for equal margins

\usepackage{geometry}
\usepackage{booktabs} % For formal tables

\usepackage{mathptmx} % This is Times font

\usepackage{fancyhdr}
\usepackage[normalem]{ulem}
\usepackage{microtype}

\usepackage{graphicx}
\usepackage{subfigure}
\graphicspath{{./figures/}}
\usepackage{url}
\usepackage{colortbl}
\definecolor{mygray}{gray}{.88}
\usepackage{multirow}
\usepackage{amsmath}
\usepackage{marvosym}
\usepackage{bm}
\usepackage{epsfig}

\usepackage{authblk}

\newcommand{\tabincell}[2]{\begin{tabular}{@{}#1@{}}#2\end{tabular}}

\begin{document} 

%%%%封面内容编辑%%%%
\begin{titlepage} % Suppresses headers and footers on the title page

	\centering % Centre everything on the title page
	
	\scshape % Use small caps for all text on the title page
	
	\vspace*{\baselineskip} % White space at the top of the page
	
	%------------------------------------------------
	%	Title
	%------------------------------------------------
	
	\rule{\textwidth}{1.6pt}\vspace*{-\baselineskip}\vspace*{2pt} % Thick horizontal rule
	\rule{\textwidth}{0.4pt} % Thin horizontal rule
	
	\vspace{0.75\baselineskip} % Whitespace above the title
	
	{\LARGE Data Dwarfs: \\A Lens Towards Fully Understanding\\Big Data and AI Workloads\\} % Title
	
	\vspace{0.75\baselineskip} % Whitespace below the title
	
	\rule{\textwidth}{0.4pt}\vspace*{-\baselineskip}\vspace{3.2pt} % Thin horizontal rule
	\rule{\textwidth}{1.6pt} % Thick horizontal rule
	
	\vspace{2\baselineskip} % Whitespace after the title block
	
	%------------------------------------------------
	%	Subtitle
	%------------------------------------------------
	
	%Subtitle here % Subtitle or further description
	
	\vspace*{3\baselineskip} % Whitespace under the subtitle
	
	%------------------------------------------------
	%	Editor(s)
	%------------------------------------------------
	
	Edited By
	
	\vspace{0.5\baselineskip} % Whitespace before the editors
	
	{\scshape\Large Wanling Gao\\ Jianfeng Zhan\\ Lei Wang\\ Chunjie Luo\\ Daoyi Zheng\\Fei Tang\\ Biwei Xie\\ Chen Zheng\\ Qiang Yang\\} % Editor list
	
	\vspace{0.5\baselineskip} % Whitespace below the editor list

	\vfill % Whitespace between editor names and publisher logo
	
	%------------------------------------------------
	%	Publisher
	%------------------------------------------------
	
	%\plogo % Publisher logo
	%\def\BUlogo{\epsfig{file=ICT.pdf,height=3cm}}
	%\includegraphics[scale=0.135]{ICT.pdf}
	\epsfig{file=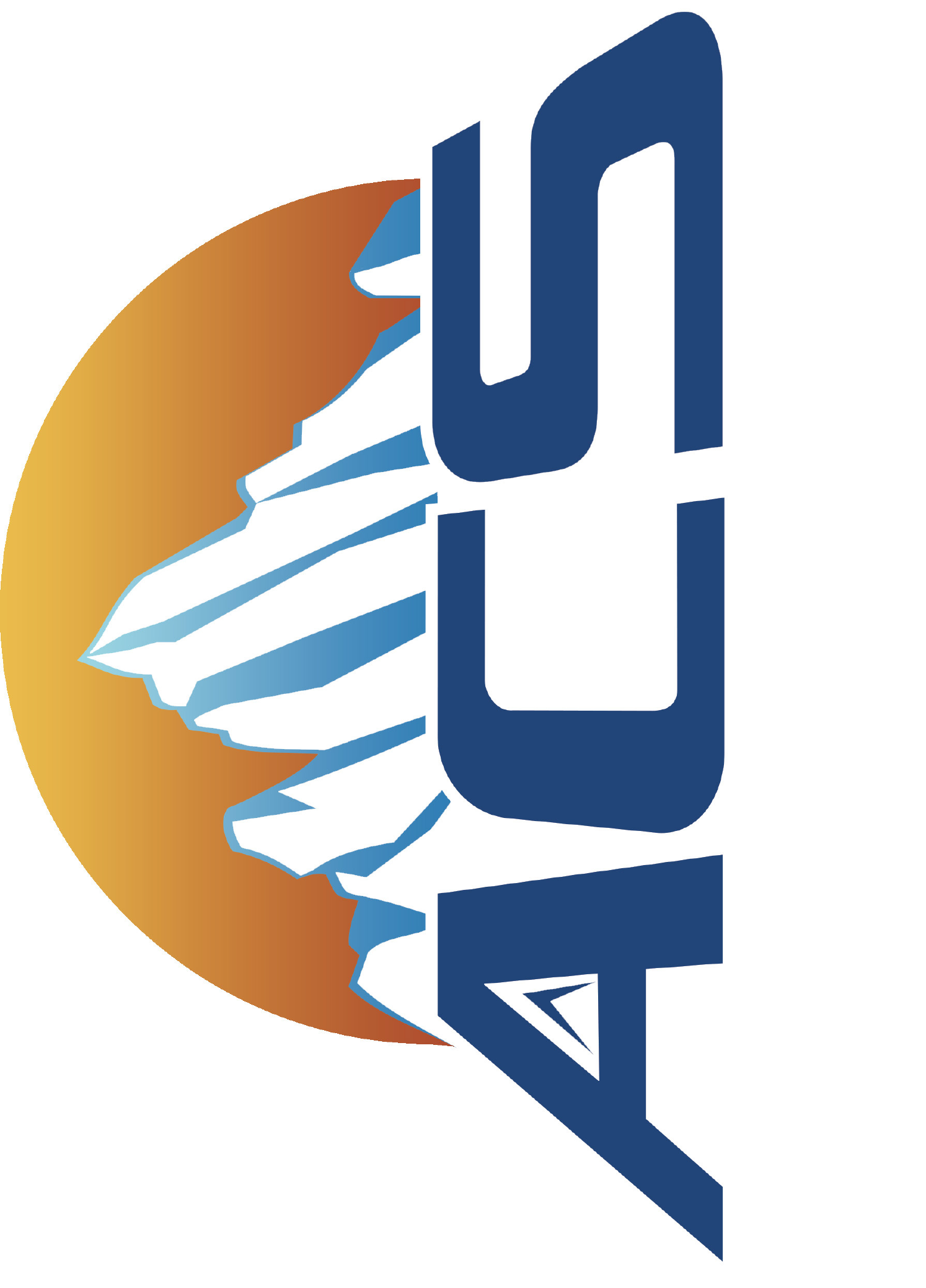,height=5cm,angle=270}
	\textit{\\Software Systems Laboratory (SSL), ACS\\ICT, Chinese Academy of Sciences\\Beijing, China\\http://prof.ict.ac.cn/ssl} % Editor affiliation
	\vspace{5\baselineskip} % Whitespace under the publisher logo

	Technical Report No. ACS/SSL-2017-3 % Publication year
	
	{\large May 3, 2018} % Publisher

\end{titlepage}

%----------------------------------------------------------------------------------------

%%%title here%%%
\title{Data Dwarfs: A Lens Towards Fully Understanding Big Data and AI Workloads}

\author{Wanling Gao, Jianfeng Zhan, Lei Wang, Chunjie Luo, Daoyi Zheng, Fei Tang, Biwei Xie, Chen Zheng and Qiang Yang}

\date{May 3, 2018}
\maketitle

\begin{abstract}

%Though  the big data benchmark suites like BigDataBench and CloudSuite have been used  in architecture and system researches, we have not yet answered the fundamental issue---what are abstractions of frequently-appearing units of computation in big data and AI workloads, which we call \emph{data dwarfs}.
%, which is the first step towards fully understanding big data workloads.

The complexity and diversity of big data and AI workloads make understanding them difficult and challenging.
This paper proposes a new approach to characterizing big data and AI workloads. We consider each big data and AI workload as a pipeline of one or more classes of unit of computations performed on different initial or intermediate data inputs. Each class of unit of computation captures the common requirements  while being reasonably divorced from individual implementations, and hence we call it a \emph{data dwarf}.
For the first time, among a wide variety of big data and AI workloads, we identify eight \emph{data dwarfs} that  takes up most of run time, including \emph{Matrix}, \emph{Sampling}, \emph{Logic}, \emph{Transform},
\emph{Set}, \emph{Graph}, \emph{Sort} and \emph{Statistic}.
%Through profiling, we found those eight dwarfs
%. different data inputs with different. As data input size increases patterns and memory access patterns, but also I/O patterns including disk I/O and network I/O patterns.
%We found combining one or more big data dwarfs with different weights can compose most of big data applications.
We implement the eight data dwarfs on different software stacks as the micro benchmarks of an open-source big data and AI benchmark suite, and perform comprehensive characterization of those data dwarfs from perspective of data sizes, types, sources, and patterns as a lens towards fully understanding big data and AI workloads.

%We analyze big data and AI workloads from the perspectives of data dwarfs and propose that using these data dwarfs instead of complex big data or AI workloads to conduct the system and architecture evaluation. Further, we present the micro-architectural characterizations of all data dwarfs and we can find that these data dwarfs can cover the diversity of big data and AI workload behaviors.

%We present the application of the dwarfs to construct big data and AI proxy benchmarks using the directed acyclic graph (DAG)-like combinations of the dwarf components with different weights to mimic the benchmarks in BigDataBench. Our proxy benchmarks shorten the execution time by 100s times on the real systems while they are qualified for both earlier architecture design and later system evaluation across different architectures.

\end{abstract}

\clearpage

\section{Introduction}

The complexity and diversity of big data and AI workloads make understanding them difficult and challenging.
First, modern big data and AI workloads expand and change very fast, and it is impossible
to create a new benchmark or proxy for every possible
workload. Second, whatever early in the architecture design process
or later in the system evaluation, it is time-consuming to run
a comprehensive benchmark suite. The complex software
stacks of the modern workloads aggravate this issue. The big data benchmark suites like BigDataBench~\cite{wang2014bigdatabench} or CloudSuite~\cite{ferdman2011clearing} are too huge to run on simulators and hence challenge time-constrained simulation and even make it impossible. Third, too complex workloads are not helpful for both reproducibility
and interpretability of performance data in benchmarking systems.

Identifying abstractions of time-consuming units of computation
is an important step toward fully understanding complex workloads.
Much previous work~\cite{codd1970relational,chen2014tpc,colella2004defining,asanovic2006landscape,shah2010data} has illustrated the importance of abstracting workloads in corresponding domains. TPC-C~\cite{chen2014tpc} is a successful benchmark built on the basis of frequently-appearing operations in the OLTP domain.
HPCC~\cite{luszczek2006hpc} adopts a similar method to design a benchmark suite for high performance computing.
Unfortunately, to the best of our knowledge, none of previous work has identified time-consuming  classes of unit of computation in big data and AI workloads.  National Research Council proposed seven major tasks in massive data analysis~\cite{council2013frontiers}, while they are macroscopical definition of problems from the perspective of mathematics.
, rather than identifying time-consuming classes of unit of computation in Big Data and AI workloads .

In this paper, we propose a new approach to characterize big data and AI workloads. We consider each big data and AI workload as a pipeline of one or more classes of   unit of computation on different initial or intermediate data inputs, each
of which captures the common requirements while being
reasonably divorced from individual implementations. We call this abstraction  a data dwarf. \emph{Significantly different from the traditional kernels, a data dwarf's behaviors are affected by the sizes, patterns, types,  and sources of different data inputs; moreover it reflects not  only computation patterns, memory access patterns, but also  disk and network I/O patterns}.

After thoroughly analyzing a majority of workloads in five
typical big data application domains (search engine, social network, e-commerce, multimedia and bioinformatics), we identify eight data dwarfs that takes up most of run time,
including \emph{Matrix}, \emph{Sampling}, \emph{Logic}, \emph{Transform},
\emph{Set}, \emph{Graph}, \emph{Sort} and \emph{Statistic}, the combinations of which describe most of big
data and AI workloads we investigated.
Considering various data inputs---text, sequence, graph, matrix and image data---with different data types and distributions,  we implement eight dwarfs on different software stacks, including Hadoop~\cite{hadoopweb}, Spark\\~\cite{zaharia2010spark}, TensorFlow~\cite{abadi2016tensorflow} and POSIX-thread (Pthread)~\cite{barney2009posix}.  For big data, the implemented data dwarfs include sort (\emph{Sort}), wordcount (\emph{Statistics}), grep (\emph{Set}), MD5 hash (\emph{Logic}), matrix multiplication (\emph{Matrix}), random sampling (\emph{Sampling}), graph traversal (\emph{Graph}) and FFT transformation (\emph{Transform}), while for AI, we implement 2-dimensional convolution (\emph{Transform}), max pooling (\emph{Sampling}), average pooling (\emph{Sampling}), ReLU activation (\emph{Logic}), sigmoid activation (\emph{Matrix}), tanh activation (\emph{Matrix}), fully connected (\emph{Matrix}), and element-wise multiplication (\emph{Matrix}), which are frequently-used computation in neural network modelling. We release the implemented data dwarfs as the micro benchmarks of an open-source big data benchmark suite. In the rest of paper, we use the big data dwarfs to indicate the dwarf implementations for big data, and use the AI dwarfs to indicate the dwarf implementations for AI.

Just like relation
algebra in database,  the data  dwarfs are promising fundamental
concepts and tools for benchmarking, designing, measuring,
and optimizing big data and AI systems. In this paper, we call attention to  performing  comprehensive characterization of those data dwarfs from perspective of data sizes, types, sources, and patterns as a lens towards fully understanding big data and AI workloads.
%focus on the application of the eight dwarfs to understand big data and AI workloads.
On a typical state-of-practice processor: Intel Xeon E5-2620 V3, we comprehensively characterize all data dwarf implementations and identify their bottlenecks.

Our contributions are five-fold as follows:

\begin{itemize}
\item We identify eight data dwarfs through  profiling  a wide variety of big data and AI workloads.
 %analysis, which capture the units of computation that perform on initial  or intermediate data input.
\item We provide diverse data dwarf implementations  on the software stacks of Hadoop, Spark, TensorFlow, Pthread.
\item From the system and micro-architecture perspectives, we comprehensively characterize the behaviors of data dwarfs and identify their bottlenecks.
%, in view of the data impacts.
We find that these data dwarfs cover a wide variety of performance space, from the perspectives of system and micro-architecture behaviors.
Moreover, the behavior of each dwarf is not only influenced by its algorithm, but also largely affected by the type, source, size, and pattern of input data.
\item From the system aspect,
we find that some AI dwarfs like convolution, fully-connected are CPU-intensive, while the other AI dwarfs are not CPU-intensive, such as Relu, Sigmoid used as activation layer.
Further, the AI dwarfs have little pressure on disk I/O, since they load a batch (e.g. 128 images) from disk every iteration.
    \item From the micro-architecture aspect, we find that these dwarfs show  various computation and memory access patterns, exploiting different parallelism degrees of ILP and MLP. With the data size expands, the percentage of frontend bound decreases while the backend bound increases.

\end{itemize}

The rest of the paper is organized as follows.
Section 2 illustrates the motivation of identifying data dwarfs.
Section 3 introduces data dwarf identification methodology. Section 4 performs system and micro-architecture evaluations on the data dwarf implementations. In Section 5, we report the data impact on the data dwarfs' behaviors from perspectives of data size, data pattern, data type and data source.
Section 6 introduces the related work. Finally, we draw a conclusion in Section 7.

\section{Motivation}

We take two examples to explain why we should call attention to performing comprehensive characterization of those data dwarfs.

%\subsection{Workloads Breakdown}
%Big data and AI dwarfs are promising fundamental tools for benchmarking, designing, measuring, and optimizing big data and AI systems. %Fig.~\ref{relation} illustrates the relationship between real workloads and eight data dwarfs.
%However, the workload quantity and fast-changing characteristics make it impossible to understand and evaluate all of them.
%In this section,
%to illustrate the importance of identifying the units of computation among a wide variety of big data and AI workloads,
%we take two important and representative big data and AI workload as examples and understand them from the algorithmic and profiling levels, respectively.
%From the algorithmic level, we understand their processing logic and abstract high-weight computation patterns; From the profiling level, we use open source tools to monitor their hotspot functions and function-call graph.
%We use Perf~\cite{}--a Linux profiling tool--to profile big data workloads, and use TensorBoard~\cite{} to visualize AI workloads.

\subsection{\textbf{SIFT Workload in Computer Vision}}

%We understand workloads using a DAG-like structure of combining one or more dwarfs. For big data and AI workloads, we explain how to use eight dwarfs to compose the original workload and illustrate their combinations using SIFT workload and AlexNet workload, respectively.

%SIFT is a typical workload for feature extraction, proposed by D. G. Lowe~\cite{lowe2004distinctive} and used to detect and describe local features of input images.
SIFT~\cite{lowe2004distinctive} is a typical workload for feature extraction, and widely used to detect local features of input images.

\begin{figure}[!t]
\centering
\includegraphics*[scale=0.6]{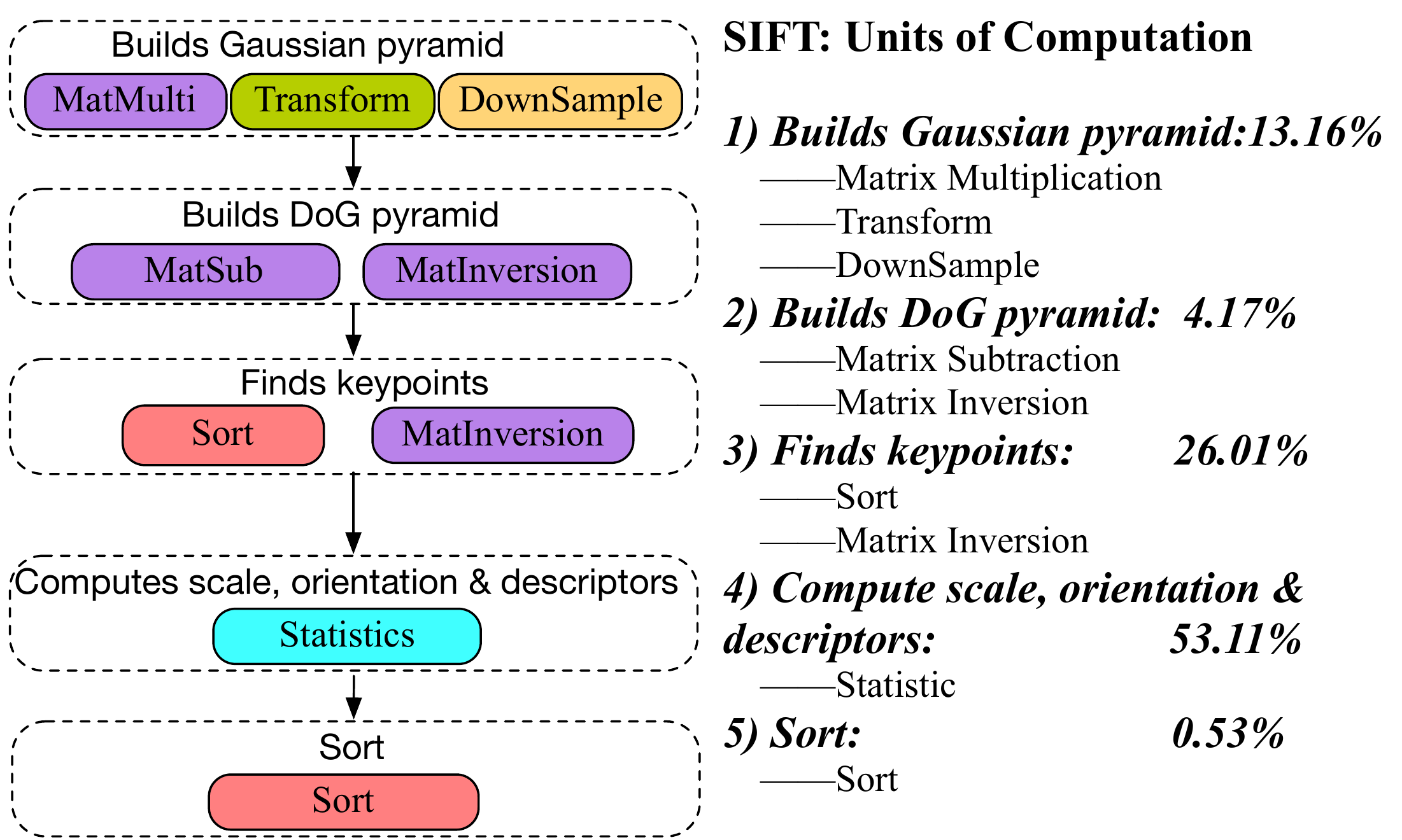}
\caption{The Computation Dependency Graph and Run Time Breakdown of SIFT Workload.} %SIFT as a representative workload in computer vision, is decomposed into several dwarfs: transform computations(FFT, IFFT), sampling computations(downsampling), matrix computations(matrix multiplication/subtraction), sort computations(sort), statistic computations(count).} %\vspace{1pt}
\label{SIFT}
\end{figure}

%We profile the SIFT workload using Perf tool.
Fig.~\ref{SIFT} shows the computation dependency graph and run time breakdown of SIFT workload. %a DAG-like structure specifies how data set or intermediate data set are operated by different dwarfs.
In total, SIFT involves five data dwarfs.
Gaussian filters $G(x,y,\partial)$ with different space scale factors $\partial$ are used to generate a group of image scale spaces, through the convolution with the input image.
%According to convolution theorem, FFT is one fast implementation method for convolution, in this regard, we don't add convolution to our list of dwarfs though it is of great significance, especially in image processing.
Image pyramid is to downsample these image scale spaces.
DOG image means difference-of-Gaussian image, which is produced by matrix subtraction of adjacent image scale spaces in image pyramid.
After that, every point in one DOG scale space would sort with eight adjacent points in the same scale space and points in adjacent two scale spaces, to find the key points in the image.
%Through computing the mold and direction of each key point and sampling in adjacent gaussian window, following by sort and statistic operations, we can get the feature vectors of the image.
Through profiling, we find that \emph{computes descirptors, finds keypoints} and \emph{builds gaussian pyramid} are three main time-consuming parts of  the SIFT workload. Furthermore, we analyze those three parts and find they are consist of  several classes of unit of computation, like Matrix, Sampling, Transform, Sort and Statistics, summing up to 83.23\% of the total SIFT run time. % proportion.

\subsection{\textbf{AlexNet in AI}}

AlexNet~\cite{krizhevsky2012imagenet} is a representative and widely-used convolutional neural network in deep learning.
In total, it has eight layers, including five convolutional layers and three fully connected layers.
%All convolutional layers use Relu as the activation function. A max pooling layer and a normalization layer follow the first and the second convolutional layers. After five convolutional layers, three fully connected layers are followed, with the first two having a Tanh layer and a dropout layer. The output of the last fully connected layer is fed to a softmax layer, to get a distribution over the class labels.

\begin{figure}[!t]
\centering
\includegraphics*[scale=0.6]{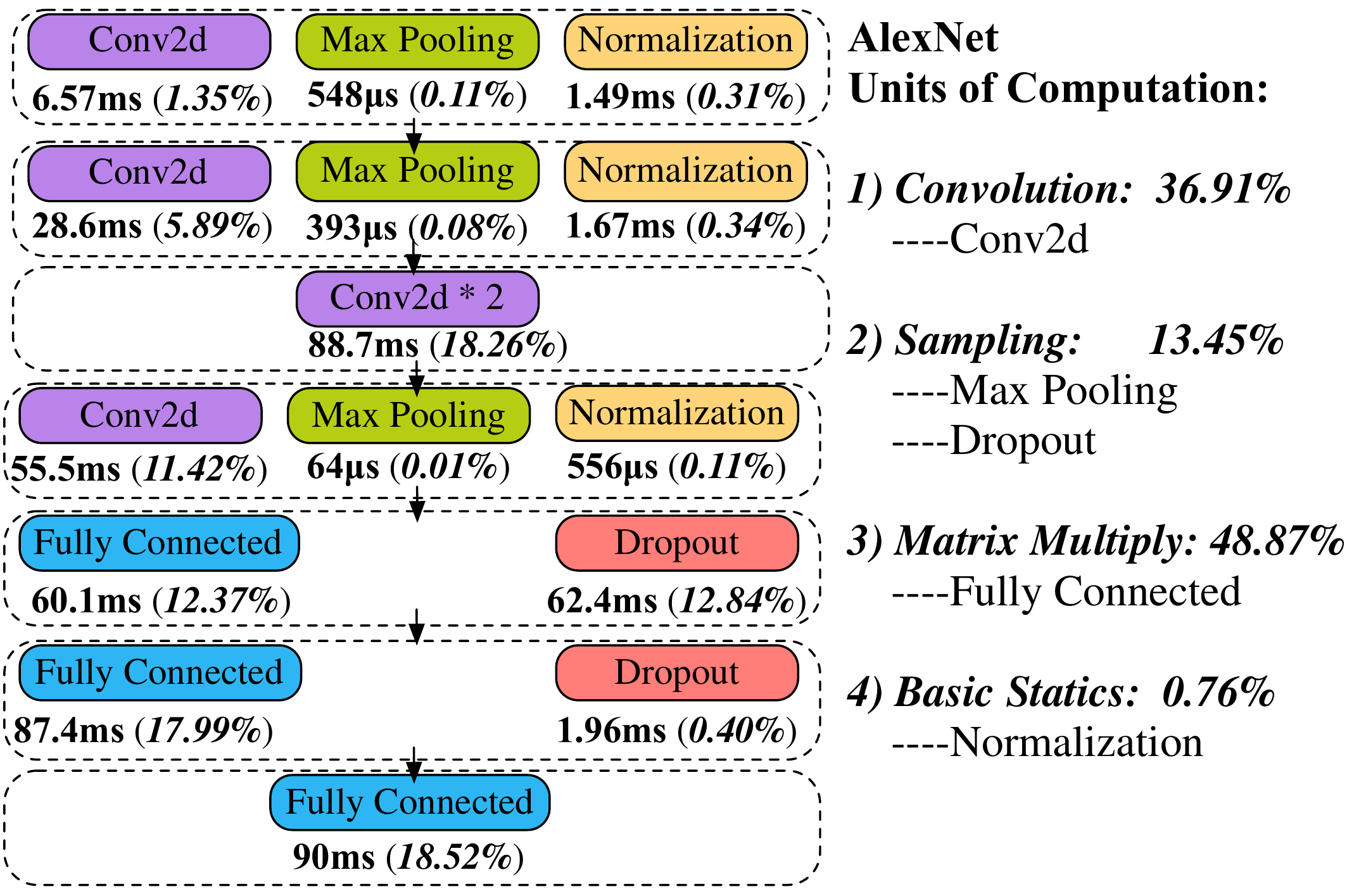}
\caption{The Computation Dependency Graph and Run Time Breakdown of One Iteration of TensorFlow AlexNet Workload.} %\vspace{1pt}
\label{alexnet}
\end{figure}

We profile one iteration of the AlexNet workload (implemented with TensorFlow) using TensorBoard toolkit and report its  computation dependency graph and run time breakdown, as shown in Fig.~\ref{alexnet}. For each operator, we report its run time and its percentage of the total run time, such as 6.57 ms and 1.35\% for the first convolution operator.
We find that each iteration involves Transform (conv2d), Sampling (max pooling, dropout),  Statistics (normalization), and Matrix (fully connected). Among them, matrix and transform computations occupy a large proportion---48.87\% and 36.91\%, respectively.

%In the next section, we will investigate a large amount of big data and AI workloads. 
Through the analysis above, we have the following observation. 
Though big data and AI workloads are very complex and fast-changing, we can consider them as a pipeline of one or more fundamental classes of unit of computation performed on different initial or intermediate data inputs.  
Those classes of unit of computation, which we call data dwarfs, occupy most of the run time of the workloads, so we should pay more attention to them.
In the next section, we will investigate more extensive big data and AI workloads, and elaborate the design of data dwarfs.

\section{Methodology}

Data dwarfs are frequently-appearing classes of unit of computation handling different data inputs.
%Different with kernels, our dwarf components consider data impacts both from data types and data sources, and have not only computation patterns and memory access patterns, but also Disk I/O patterns and network I/O patterns.
In this section, we illustrate how to identify data dwarfs from big data and AI workloads, and illustrate our data dwarf implementations.

\subsection{Dwarf Identification Methodology}

%In this section, we illustrate the big data dwarfs and corresponding dwarf component implementations.

\begin{figure}[!t]
\centering
\includegraphics*[scale=1.2]{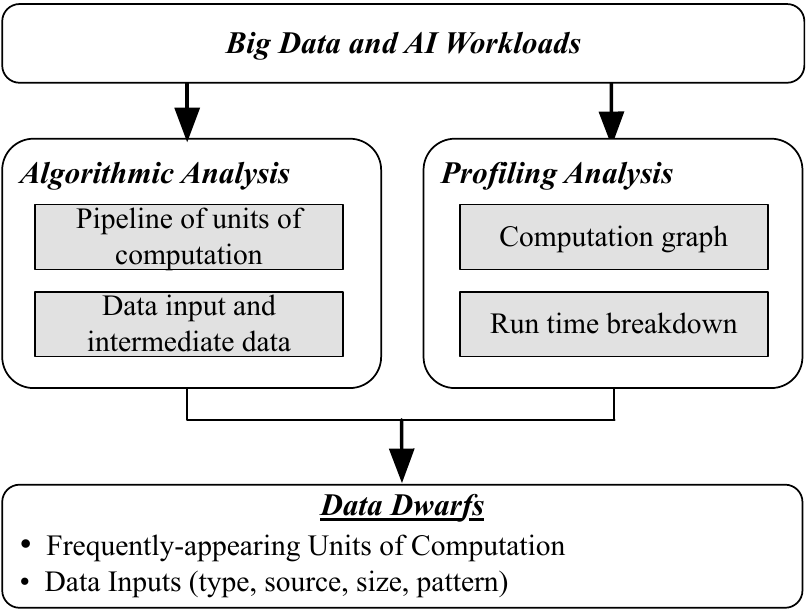}
\caption{Identifying Data Dwarfs.} %\vspace{1pt}
\label{fig_dwarf_overview}
\end{figure}

Fig.~\ref{fig_dwarf_overview} overviews the methodology of dwarf identification.
We first single out a broad spectrum of big data and AI workloads through investigating  five typical application domains (search engine, social network, e-commerce, multimedia, and bioinformatics) and representative algorithms in four processing techniques (machine learning, data mining, computer vision and natural language processing).
Then we analyze and profile these workloads. 
On one hand, we decompose the algorithm into a pipeline of units of computations and focus on the input/intermediate data as well.
On the other hand, we profile the workload to analyze the computation dependency graph and run time breakdown.
%We investigate typical algorithms in five application domains (search engine, social network, e-commerce, multimedia, and bioinformatics) and single out important algorithms widely used in big data computing.   %forty algorithms widely used in big data computing, which consist units of computation that frequently appeared in big data analytics.
%Combing algorithmic analysis and experimental analysis, we finalize eight dwarfs according to their frequency and importance, which are abstraction of frequently-appearing operations.

%After singling out a broad spectrum of big data analytics workloads (machine learning, data mining, computer vision and natural language processing) through investigating five application domains (search engine, social network, e-commerce, multimedia, and bioinformatics), we analyze these workloads and decompose them to multiple classes of units of computation.
According to the units of computation pipeline and run time breakdown, we finalize eight big data and AI dwarfs, which are essential computations that take up most of run time.
Table~\ref{statistical} shows the importance of eight data dwarfs in a majority of big data and AI workloads.
%We can find that these eight dwarfs are major classes of units of computations in a variety of big data and AI workloads.
Note that previous work~\cite{guinard2010resource} has identified four basic units of computation in online service, including get, put, post, delete. We don't include those four in our dwarf set. % for simplicity.

%\doublerulesep 0.1pt
\begin{table*}[htb]
\caption{The Importance of Eight Data Dwarfs in Big Data and AI workloads.}\label{statistical}
\renewcommand\arraystretch{1.2}
\center
\footnotesize
%\begin{footnotesize}
\begin{tabular}{|p{1in}|p{1.2in}|p{2in}|p{1.5in}|}
  \hline
  \textbf{Catergory}   & \textbf{ Application Domain } &  \textbf{Workload } &  \textbf{ Unit of Computation } \\
  \hline
  \multirow{2}{*}{\tabincell{l}{Deep Learning}} &  \multirow{2}{*}{\tabincell{l}{ Image Recognition \\ Speech Recognition }} & Convolutional neural network(CNN)  &  Matrix, Sampling, Transform \\  \cline{3-4}
                                                &  & Deep belief network(DBN)         & Matrix, Sampling \\ \cline{3-4}
  \hline

  \multirow{2}{*}{\tabincell{l}{Graph Mining}}  &  \multirow{2}{*}{\tabincell{l}{ Search Engine \\ Community Detection }}  & PageRank  & Matrix, Graph, Sort \\  \cline{3-4}
                                                &   & BFS, Connected component(CC)      &  Graph \\ \cline{3-4}
  \hline

  \multirow{2}{*}{\tabincell{l}{Dimension Reduction}}  &  \multirow{2}{*}{\tabincell{l}{ Image Processing \\ Text Processing }}  & Principal components analysis(PCA)  & Matrix \\  \cline{3-4}
                                                &   & Latent dirichlet allocation(LDA)  &  Statistics, Sampling \\ \cline{3-4}
  \hline

  \multirow{3}{*}{\tabincell{l}{Recommendation}} & \multirow{3}{*}{\tabincell{l}{Association Rules Mining \\ Electronic Commerce }}  & Aporiori  &  Statistics, Set \\  \cline{3-4}
                                                             &  & FP-Growth &  Graph, Set, Statistics \\ \cline{3-4}
                                                             &  & Collaborative filtering(CF) &  Graph, Matrix \\ \cline{3-4}
  \hline

  \multirow{4}{*}{\tabincell{l}{Classification}} &  \multirow{4}{*}{\tabincell{l}{ Image Recognition \\ Speech Recognition \\ Text Recognition }} & Support vector machine(SVM)  &  Matrix \\  \cline{3-4}
                                                 &  & K-nearest neighbors(KNN) &  Matrix, Sort, Statistics \\ \cline{3-4}
                                                 &  & Naive bayes  &  Statistic \\ \cline{3-4}
                                                 &  & Random forest  &  Graph, Statistics  \\ \cline{3-4}
                                                 &  & Decision tree(C4.5/CART/ID3)  &  Graph, Statistics  \\ \cline{3-4}
  \hline

  \multirow{1}{*}{\tabincell{l}{Clustering}} & \multirow{1}{*}{\tabincell{l}{ Data Mining }} &  K-means  & Matrix, Sort  \\ \cline{3-4}
  \hline

  \multirow{4}{*}{\tabincell{l}{Feature Preprocess}} & \multirow{4}{*}{\tabincell{l}{ Image Processing \\ Signal Processing \\ Text Processing}}  & Image segmentation(GrabCut)  & Matrix, Graph \\  \cline{3-4}
                                                     &  & Scale-invariant feature transform(SIFT)  & Matrix, Transform, Sampling, Sort, Statistics \\  \cline{3-4}
                                                     &  & Image Transform  & Matrix, Transform \\  \cline{3-4}
                                                     &  & Term Frequency-inverse document frequency (TF-IDF)   & Statistics \\  \cline{3-4}
  \hline

  \multirow{2}{*}{\tabincell{l}{Sequence Tagging}} &\multirow{2}{*}{\tabincell{l}{ Bioinformatics \\ Language Processing }}  & Hidden Markov Model(HMM)  &  Matrix \\  \cline{3-4}
                                               &  & Conditional random fields(CRF) & Matrix, Sampling \\ \cline{3-4}
  \hline

%  \multirow{2}{*}{\tabincell{l}{Hashing}} & \multirow{2}{*}{\tabincell{l}{ Cryptography \\ Digital Signature }}  & SimHash, MinHash  & Set, Logic \\  \cline{3-4}
%%                                       &  & MinHash  & Set, Logic \\  \cline{3-4}
%                                       &  & Locality-sensitive hashing(LSH)  & Set, Logic \\  \cline{3-4}
%  \hline

%  \multirow{2}{*}{\tabincell{l}{Indexing}} & \multirow{2}{*}{\tabincell{l}{ Search Engine \\ Information Retrieval }}  & Inverted index  & Basic Statistic, Logic, Set \\  %\cline{3-4}
  Indexing & Search Engine & Inverted index, Forward index & Statistics, Logic, Set, Sort\\
%                                       &  & Jaccard index  & Set \\  \cline{3-4}
  \hline

  \multirow{4}{*}{\tabincell{l}{Encoding/Decoding}} & \multirow{4}{*}{\tabincell{l}{ Multimedia Processing \\ Security \\ Cryptography \\ Digital Signature }}  & MPEG-2  & Matrix, Transform \\  \cline{3-4}
                                                    &   & Encryption   & Matrix, Logic \\  \cline{3-4}
                                                    &   & SimHash, MinHash  & Set, Logic \\  \cline{3-4}
                                                    &  & Locality-sensitive hashing(LSH)  & Set, Logic \\ \cline{3-4}
  \hline

  \multirow{1}{*}{\tabincell{l}{Data Warehouse}} & \multirow{1}{*}{\tabincell{l}{  Business intelligence }}  &  Project, Filter, OrderBy, Union  &  Set, Sort \\  \cline{3-4}
  \hline

%  \multirow{2}{*}{\tabincell{l}{Basic }} & \multirow{2}{*}{\tabincell{l}{ Search Engine \\ Text processing }}  & Wordcount  & Basic Statistic \\  \cline{3-4}
%                                         &   & TeraSort   &  Sort \\  \cline{3-4}
%                                         &   & Term Frequency-inverse document frequency (TF-IDF)   &  Basic Statistic \\  \cline{3-4}
%	  \hline

\end{tabular}
%\end{footnotesize}
\end{table*}

%\begin{figure*}[!t]
%\centering
%\includegraphics*[scale=0.68]{figures/methodology.pdf}
%\caption{Methodology of Data Dwarfs.} %\vspace{1pt}
%\label{fig:methodology}
%\end{figure*}

\subsection{Eight Data Dwarfs} \label{bigdatadwarf}

In this subsection, we summarize eight data dwarfs frequently appearing in big data and AI workloads.

\textbf{Matrix} In big data and AI workloads, many problems involve matrix computations, such as matrix multiplication and matrix transposition.

\textbf{Sampling} Sampling plays an essential role in big data and AI processing, which obtain an approximate solution when one problem cannot be solved by using analytical method.

\textbf{Logic} We name computations performing bit manipulation as logic computations, such as hash, data compression and encryption.

\textbf{Transform} The transform computations here mean the conversion from the original domain (such as time) to another domain (such as frequency). Common transform computations include discrete fourier transform (DFT), discrete cosine transform (DCT) and wavelet transform.

\textbf{Set} In mathematics, set means a collection of distinct objects. Likewise, the concept of set is also widely used in computer science.
For example, similarity analysis of two data sets involves set computations, such as Jaccard similarity. Furthermore, fuzzy set and rough set play very important roles in computer science.

\textbf{Graph} A lot of applications involve graphs,
with nodes representing entities and edges representing dependencies.
Graph computation is notorious for having irregular memory access patterns.

\textbf{Sort} Sort is widely used in many areas. Jim Gray thought sort is the core of modern databases~\cite{asanovic2006landscape}, which shows its fundamentality.

\textbf{Statistics} Statistic computations are used to obtain the summary information through statistical computations, such as counting and probability statistics.

\subsection{Data Dwarf Implementations} \label{component}

Data dwarfs are the fundamental components of big data and AI workloads, which is of great significance for evaluation, considering the complexity and diversity of big data and AI workloads.
We provide the data dwarf implementations for big data and AI separately, according to their computation specialties.
%For multi-dimensional evaluation, i.e. system, micro-architecture, data management, we implement data dwarfs with various software stacks.
For the big data dwarf implementations, we provide Hadoop~\cite{hadoopweb}, Spark~\cite{zaharia2010spark}, and Pthreads~\cite{barney2009posix} implementations. %for data dwarfs widely used in big data processing. 
These data dwarfs include sort, wordcount, grep, MD5 hash, matrix multiplication, random sampling, graph traversal and FFT transformation.
For the AI dwarfs, we provide TensorFlow~\cite{abadi2016tensorflow} and Pthread implementations, 
 %for particular data dwarfs widely used in artificial intelligence, 
 including 2-dimensional convolution, max pooling, average pooling, relu activation, sigmoid activation, tanh activation, fully connected (matmul), and element-wise multiply.
 %, according to the common and important layers in a neural network structure.
We consider the impact of data input from the perspectives of type, source, size, and pattern. Among them, \emph{data type} includes structure, un-structured, and semi-structured data. \emph{Data source} indicates the data storage format, including text, sequence, graph, matrix, and image data. \emph{Data pattern} includes the data distribution, data sparsity. As for \emph{data size}, we provide big data generators for text, sequence, graph and matrix data to fulfill different size requirements.

\section{Characterization}\label{evaluation}

%In this section, we evaluate our dwarf-based simulation version for three representative Hadoop-based big data analytics workloads, from the perspective of simulation time, simulation accuracy and data Adaptability.

In this section, we evaluate data dwarfs with various software stacks from the perspectives of both system and architecture behaviors.

\subsection{Experiment Setups}

%To verify effectiveness of our dwarf-based simulation method, we verify the similarity of three representative Hadoop-based workloads and corresponding dwarf-based workloads.
%The chosen three representative big data analytics workloads are Hadoop Terasort (CPU-intensive at Map stage and I/O-intensive at Reduce stage), Hadoop Kmeans (CPU-intensive) and Hadoop PageRank (Hybrid).

We deploy a three-node cluster, with one master node and two slave nodes.
They are connected using 1Gb Ethernet network.
Each node is equipped with two Intel Xeon E5-2620 V3 (Haswell) processors, and each processor has six physical out-of-order cores.
The memory of each node is 64 GB.
%Each node is equipped with two Intel Xeon Er-2620 V3 (Haswell) processors, which consist of six physical out-of-order cores, and 64GB DDR4 memory.
The operating system, software stacks and gcc versions are as follows: CentOS 7.2 (with kernel 4.1.13); JDK 1.8.0\_65; Hadoop 2.7.1; Spark 1.5.2; tensorFlow 1.0; GCC 4.8.5.
%Each node runs Linux CentOS 7.2 with the Linux kernel version 4.1.13. The JDK and Hadoop version is  1.8.0\_65 and 2.7.1, respectively.
%The Spark and TensorFlow version is 1.5.2 and 1.0, respectively. The GCC version is 4.8.5.
The data dwarfs implemented with Pthread are compiled using "-O2" option for optimization.
The hardware and software details are listed in Table \ref{hwconfigeration}.
Since Pthread is a multi-thread programming model, we evaluate both the TensorFlow and Pthread implementations of AI dwarfs on one node for apple-to-apple comparison.

%To evaluate the performance data accuracy, we run the proxy benchmarks against the  benchmarks from BigDataBench. We run the four Hadoop benchmarks from BigDataBench on the above five-node cluster using the optimized Hadoop configurations, through tuning the data block size of the Hadoop distributed file system, memory allocation for each map/reduce job and reduce job numbers according to the cluster scale and memory size.
%For Hadoop TeraSort, we choose 100 GB text data produced by gensort~\cite{gensort}. For Hadoop Kmeans and PageRank, we choose 100 GB sparse vector data with 90\% sparsity~\footnote{The sparsity of the vector indicates the proportion of zero-valued elements.} and $2^{26}$-vertex graph both generated by BDGS~\cite{ming2014bdgs}, respectively. For Hadoop SIFT, we use one hundred thousand images from ImageNet~\cite{imagenet_cvpr09}.
%The dwarf benchmarks are tuned through our simulation methodology, with the input data size satisfying the time constraint.
%For comparison, we run the four proxy benchmarks on one of the slave nodes, respectively.

% \textbf{\emph{{how about the dwarf version?)}}}

\begin{table}
\caption{Configuration Details of Xeon E5-2620 V3}\label{hwconfigeration}
\renewcommand\arraystretch{1.2}
\center
\footnotesize
\begin{tabular}{|p{0.8in}|p{0.8in}|p{0.8in}|p{0.8in}|}
\hline \rowcolor{mygray} \multicolumn{4}{|l|}{Hardware Configurations}\\
\hline \multicolumn{2}{|c|}{CPU Type} & \multicolumn{2}{c|}{Intel CPU Core} \\
\hline \multicolumn{2}{|c|}{Intel \textregistered Xeon E5-2620 V3}  &\multicolumn{2}{c|}{12 cores@2.40G} \\
\hline L1 DCache &L1 ICache &L2 Cache &L3 Cache \\
\hline 12 $\times$ 32 KB& 12 $\times$ 32 KB&12 $\times$ 256 KB& 15MB \\
\hline \multicolumn{2}{|c|}{Memory} & \multicolumn{2}{c|}{64GB,DDR4}  \\
\hline \multicolumn{2}{|c|}{Disk} & \multicolumn{2}{c|}{SATA@7200RPM}\\
\hline \multicolumn{2}{|c|}{Ethernet} & \multicolumn{2}{c|}{1Gb}\\
\hline \multicolumn{2}{|c|}{Hyper-Threading} & \multicolumn{2}{c|}{Disabled}\\
%\hline
%\hline \rowcolor{mygray} \multicolumn{4}{|l|}{Software Configurations}\\
%\hline \multicolumn{2}{|c|}{Operating System} & \multicolumn{2}{c|}{CentOS 7.2}  \\
%\hline \multicolumn{2}{|c|}{Linux Kernel} & \multicolumn{2}{c|}{4.1.13}  \\
%\hline \multicolumn{2}{|c|}{JDK Version} & \multicolumn{2}{c|}{1.8.0\_65}  \\
%\hline \multicolumn{2}{|c|}{Hadoop Version} & \multicolumn{2}{c|}{2.7.1}  \\
%\hline \multicolumn{2}{|c|}{Spark Version} & \multicolumn{2}{c|}{1.5.2}  \\
%\hline \multicolumn{2}{|c|}{Tensorflow Version} & \multicolumn{2}{c|}{1.0}  \\
%%\hline \tabincell{l}{Operating\\System} & \tabincell{l}{Linux\\Kernel} & \tabincell{l}{JDK\\Version} & \tabincell{l}{Hadoop\\Version} \\
%%\hline CentOS 7.2 & 4.1.13 & 1.8.0_65 & 2.7.1\\
\hline
\end{tabular}
\end{table}

\subsection{Experiment Methodology}

We evaluate eight big data dwarfs implemented with Hadoop, Spark, and eight AI data dwarfs implemented with TensorFlow and Pthread.
Note that we use the optimal configurations for each software stack, according to the cluster scale and memory size.
The data configuration and selected metrics are listed as follows.

%To evaluate accuracy, we choose  micro-architectural and system metrics covering instruction mix, cache behavior, branch prediction, processor performance, memory bandwidth and disk I/O behavior.
\textbf{Data Configuration}
To evaluate the impacts of data input comprehensively, we evaluate the data dwarfs with three data sizes: \emph{Small}, \emph{Medium}, and \emph{Large}.
For the graph dwarf, \emph{Small}, \emph{Medium}, \emph{Large} is $2^{22}$, $2^{24}$ and $2^{26}$-vertex, respectively. For the matrix dwarf, we use 100, 1K and 10K two-dimensional matrix data with the same distribution and sparsity. For the transform dwarf, we use 16384, 32768 and 65536 two-dimension matrix data. For the other big data dwarfs, we use 1, 10 and 100 GB wikipedia text data, respectively.
For the AI dwarfs, we use three configurations in terms of input tensor sizes and channels. They are \emph{(224*224,64), (112*112,128) and (56*56,256)}. Among them, the first value indicates the dimension of input tensor, the second value indicates the channels, and  all of them use 128 as batch size. We choose these three configurations because they are widely used in neural network models~\cite{simonyan2014very}. Note that the dimension for all input tensors is 224 for \emph{Large} configuration, 112 for \emph{Medium} configuration and 56 for \emph{Small} configuration.
For the Pthread-version AI dwarfs, we use 1K, 10K, 100K images from ImageNet~\cite{deng2009imagenet}.
%In this section, we only report the evaluation on the data dwarfs with the \emph{Large} configuration. In the next section (Section~\ref{dataimpact}), we will report the data impacts using the above three-size configurations.
In the following sections, we characterize the system and micro-architectural behaviours of data dwarfs with the \emph{Large} data size.
In Section~\ref{dataimpact}, we will analyze the impact of data input on characteristics with all data sizes.

%For the eight big data dwarfs, we use $2^{26}$-vertex graph data for graph dwarfs, and 10000*10000 matrix data for matrix dwarfs. For transform dwarf, we use 65536*65536 dense matrix data. For the other dwarfs, we use 100 GB wikipedia text data.

%For TensorFlow workloads, we use the 224*224 dimension tensors with 64 channels, and 128 as the batch size, which widely used in neural network modelling~\cite{simonyan2014very}.

\textbf{System and Micro-architecture Metrics}
%We characterize the system and micro-architectural behaviors of the data dwarfs.
We characterize the system and micro-architectural behaviors~\cite{van2016analytical} of the data dwarfs, which is significant for design and optimization~\cite{quinn2015using}.
For system evaluation, we report the metrics of CPU utilization, I/O Wait, disk I/O bandwidth and, network I/O bandwidth. The system metrics are collected through the proc file system.

For micro-architecture evaluation, we use the Top-Down method~\cite{yasin2014top}, which categorizes the pipeline slots into four categories, including retiring, bad speculation, frontend bound and backend bound. Among them, retiring represents the useful work, which means the issued micro operations (uops) eventually get retired. Bad speculation represents the pipeline is blocked due to incorrect speculations. Frontend bound represents the stalls due to frontend, which undersupplies uops to the backend. Backend bound represents the stalls due to backend, which is a lack of required resources for new uops~\cite{pmutools}.
We use Perf~\cite{perftool}, a Linux profiling tool, to collect the hardware events referring to the Intel Developer\'s Manual~\cite{guide2011intel} and pmu-tools~\cite{pmutools}.

\subsection{System Evaluation}\label{system:exp}

\begin{figure*}[!t]
\centering
\includegraphics*[scale=0.6]{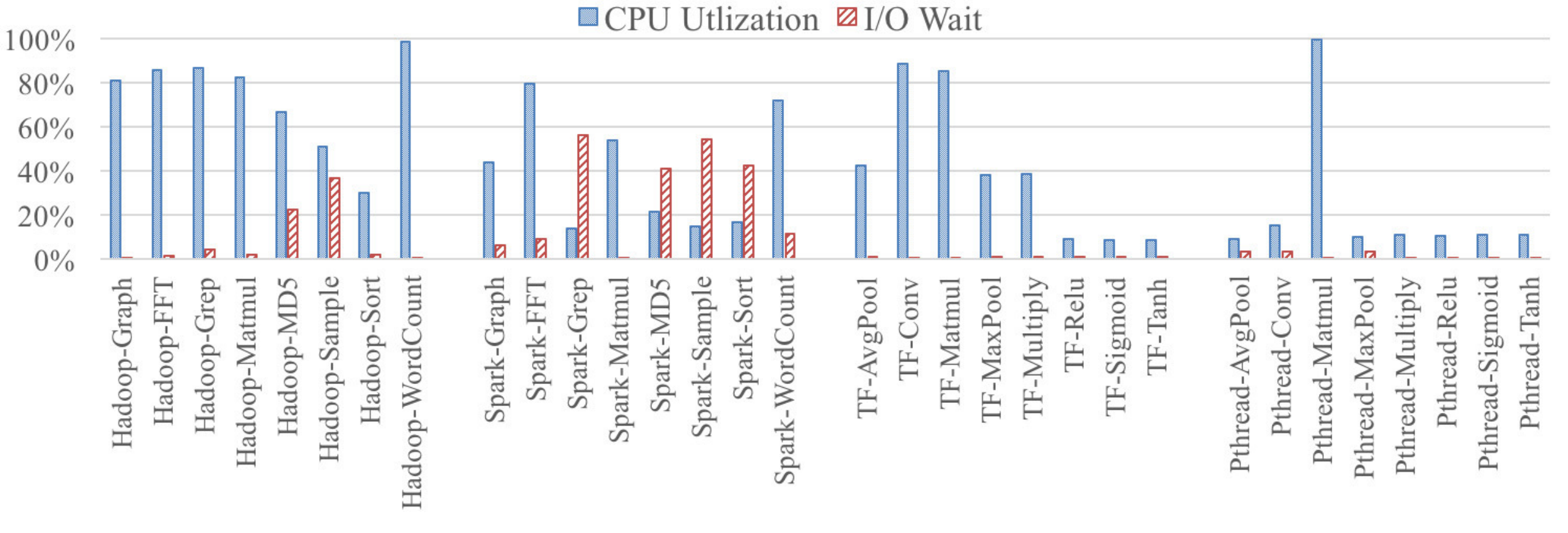}
\caption{CPU Utilization and I/O Wait of Data Dwarfs.} %\vspace{1pt}
\label{cpuiowait}
\end{figure*}

\begin{figure}[!t]
\centering
\includegraphics*[scale=0.9]{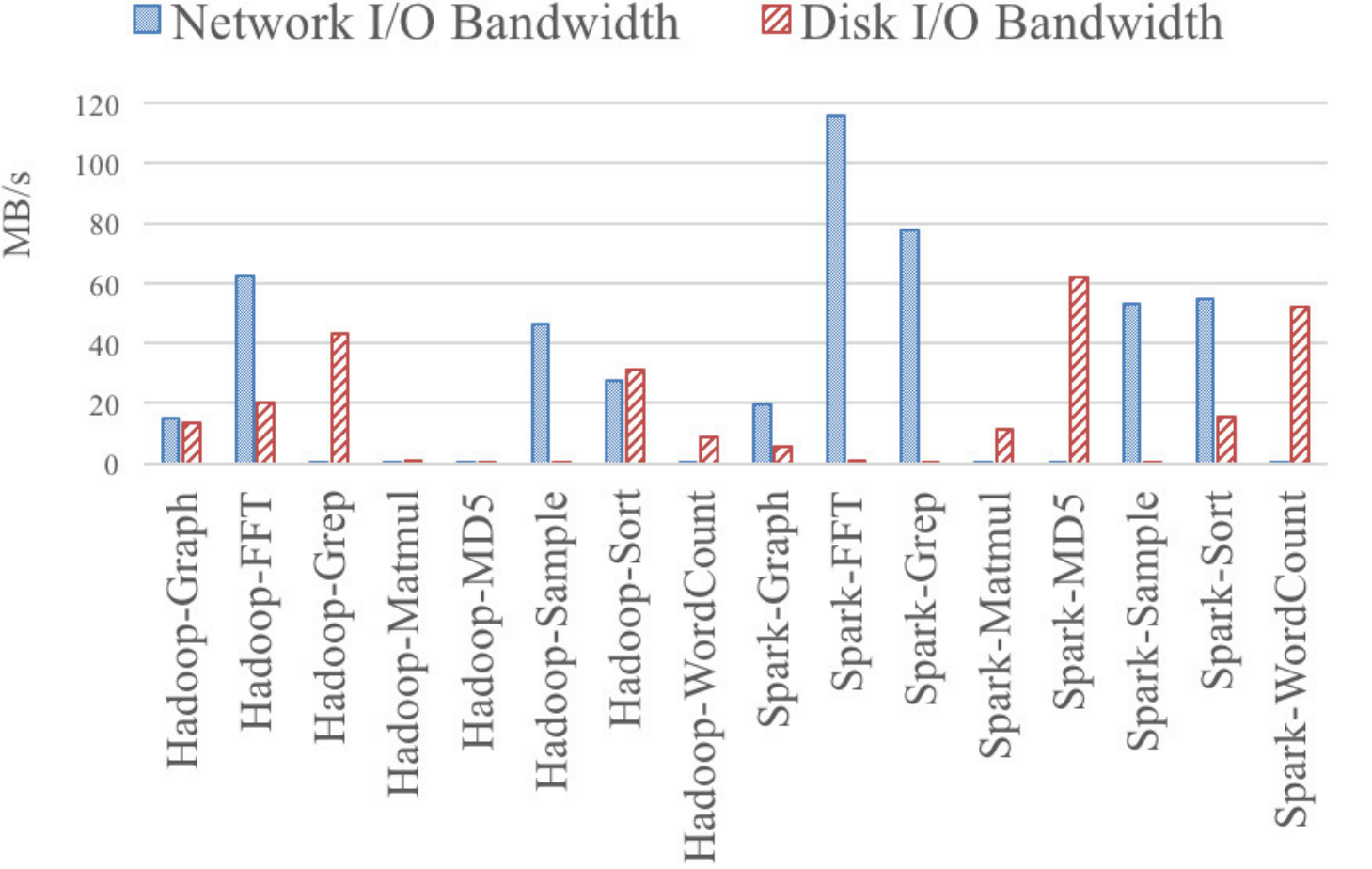}
\caption{I/O Behaviors of Data Dwarfs.} %\vspace{1pt}
\label{io}
\end{figure}

Fig.~\ref{cpuiowait} presents the CPU utilization and I/O Wait of all data dwarfs. We find that Hadoop dwarfs have higher CPU utilization than Spark dwarfs, and suffers less I/O Wait than Spark dwarfs do. Particularly, Hadoop dwarfs  take 80 percent CPU time.
The I/O Waits of AI data dwarfs are extremely lower than that of big data dwarfs.
For deep neural networks, even the total input data is large, the input layer loads a batch from disk every iteration, so data loading size from disk by the input layer occupies a very small proportion comparing to intermediate data, and thus introduce  little disk I/O requests.
Pthread dwarfs has less CPU utilization and I/O Wait in general, because Pthreads dwarfs have less memory allocation and relocation operations than counterparts using other stacks. Moreover, the data loading time overlaps the processing time since computation is simple, except that Pthread Matmul has almost 100\% CPU utilization because it is very CPU-intensive.
 %computation is very complexcomplexity.
Tensorflow dwarfs, such as AvgPool, Conv, Matmul, Maxpool, and Multiply, have taken most of CPU time, because these five dwarfs are CPU-intensive. Nevertheless, we also find that the other  AI dwarfs are not that CPU-intensive, such as Relu, Sigmoid, and Tanh. %In general, as the run time breakdown shown in Fig.~\ref{alexnet}, the CPU-intensive dwarfs take most part of real neural networks. As a result, AI workloads in general are CPU-intensive.

%Fig~\ref{intrctx} presents the frequency of interrupts and context switches during dwarfs execution. We find that Big Data Dwarfs have higher interrupt and context switch frequencies, while AI dwarfs has lower and similar frequencies.
%of both activities for each dwarf implementations.
%As Tensorflow and Pthread stacks have long-standing processes and static process number, they only changes when inter-process communication occurs. Hadoop and Spark are based on large amounts of short-time java processes, which make the context-switches frequency higher.
%For interrupts, AI dwarfs mainly suffer from timer interrupts, while Hadoop and Spark have much more device interrupts including disk and network I/O.

%Fig~\ref{fig:pagefault} presents the frequency of page fault and major page fault, respectively.
%We find that, big data dwarfs, especially Hadoop dwarfs have much more page faults than AI dwarfs. Hadoop stacks need to invoke large amounts of java processes, and shuffle data before merge.
%As a result, Hadoop need to remap or reallocate large amounts of memory pages, which may make performance suffer.
%Major page fault occurs only when it needs to swap data into memory, which consumes longer time than normal page fault.
%So compared to Spark FFT and Pthread MaxPool, Tensorflow dwarfs have more major page faults than other dwarfs stacks.

Fig~\ref{io} presents the network bandwidth and disk I/O bandwidth. For AI dwarfs, most
of them (e.g. matmul, relu, pooling, activation)  are executed in the hidden layers, and the intermediate states of hidden layers are stored in the memory. That is to say, the hidden layers consume the most resources of computation and memory storage, while the disk I/O for input layer is relatively minor. Our evaluation confirms this observation. Meanwhile, we mentioned in Section 4.1, we evaluate  both the TensorFlow and
Pthread implementations of AI dwarfs on one node for
apple-to-apple comparison. So we do not report the I/O behaviors of AI dwarfs.
We find that for all big data dwarfs, Spark stack has much larger network I/O pressure than that of Hadoop stack, because Spark stack has more data shuffles, so it needs transferring data from one node to another one frequently.
Five of the eight Spark implementations have smaller disk I/O pressure than that of Hadoop, because Spark targets in-memory computing.
Except Spark Matmul, Spark MD5 and WordCount have larger disk I/O pressure than that of Hadoop counterparts.
The disk I/O read sector numbers are nearly equal, while the write sector numbers are much larger.

%We find that, Big data dwarfs have much more bandwidth than AI dwarfs in both network and disk I/O.
%It is because Hadoop and Spark has a large amount of intermediate data that needed to be shared among tasks in different nodes.

\subsection{Micro-architecture Evaluation}

\begin{figure}[!t]
\centering
\includegraphics*[scale=0.75]{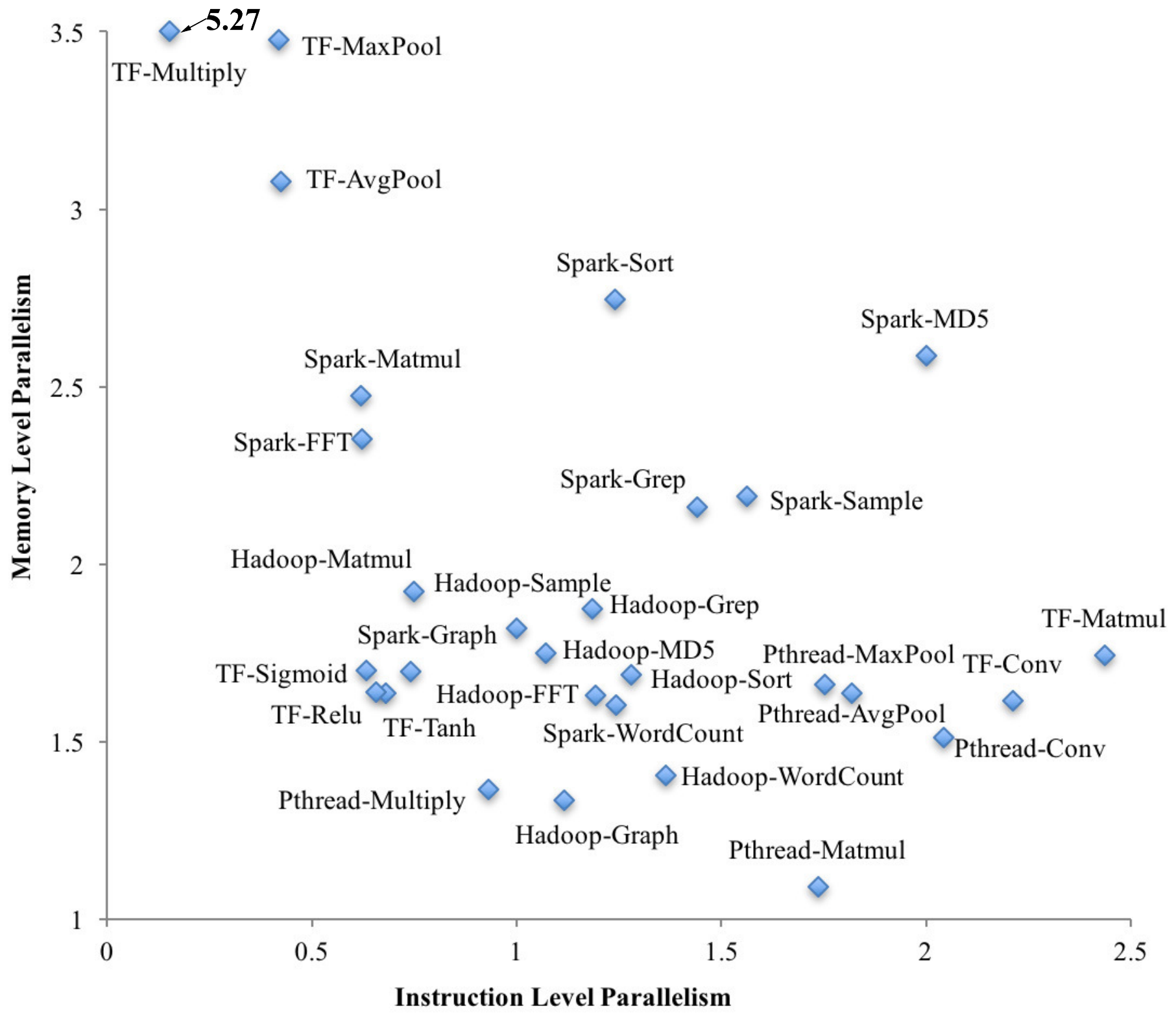}
\caption{Execution Performance of Data Dwarfs.} %\vspace{1pt}
\label{ilpmlp}
\end{figure}

To better understand the data dwarfs, we analyze their performance and micro-architectural characteristics.
%, including their frontend and backend behaviours.

\textbf{Execution Performance} The execution performance indicates the overall running efficiency of the workloads~\cite{kim2016automatically}. We use the instruction level parallelism (ILP) and memory level parallelism (MLP) to reflect the execution performance. Among them, ILP measures the number of instructions that can be executed simultaneously. Here we use the retired instructions per cycle (IPC) to measure ILP. MLP indicates the parallelism degree that memory accesses can be generated and executed~\cite{glew1998mlp}.
Fig.~\ref{ilpmlp} presents the ILP and MLP of all data dwarfs. We find that these dwarfs cover a wide range of ILP and MLP behaviors, reflecting distinct computation and memory access patterns.
For example, TensorFlow Multiply does element-wise multiplications and has high MLP (5.27) but extremely low ILP (0.15).  This is because that its computation is simple and has little data dependencies, so it generates a large amount of memory access requests while has  no enough independent instructions to execute, thus incurs severe backend stalls and results in low ILP. Also, max pooling and average pooling have high MLP.
%The reason why the MLP of average pooling lower than max pooling is that average computation involves many divide operations, and thus suffers more stalls due to the delay of divider unit.
The MLP of average pooling is lower than max pooling, because average computation involves many divide operations, and thus suffers more stalls due to the delay of divider unit.
The software stack changes workload's computation and memory access patterns, which is also found in previous work~\cite{jia_bigDataBench_subset}. For example, both Hadoop FFT and Spark FFT are based on cooley-tukey algorithm~\cite{cooley1965algorithm}, while they have different parallelism degrees. Spark FFT is more memory-intensive and has higher MLP.
%TensorFLow Matmul which uses as the fully connected layer has high ILP.
%For AI data dwarfs implemented with TensorFlow, almost all of them have low ILP but high MLP, this is because that the AI dwarf computation is concise, so the impact of TensorFlow framework and data read/write are more obvious.

\begin{figure}[!t]
\centering
\includegraphics*[scale=0.9]{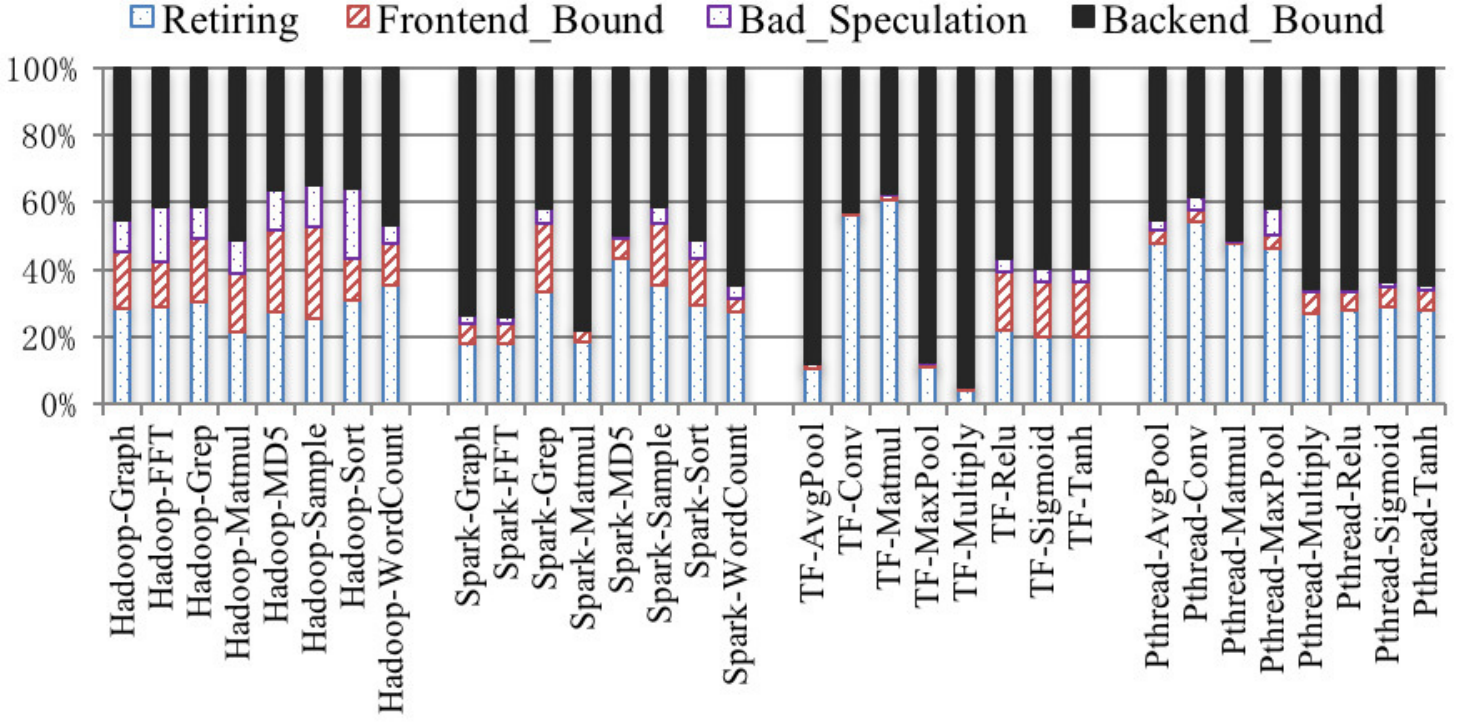}
\caption{The Uppermost Level Breakdown of Data Dwarfs.} %\vspace{1pt}
\label{level1}
\end{figure}

\textbf{The Uppermost Level Breakdown} Fig.~\ref{level1} shows the uppermost level breakdown of all data dwarfs we evaluated. We find that these dwarfs have different pipeline bottlenecks. For Hadoop dwarfs, they suffer from notable stalls due to frontend bound and bad speculation.
Moreover, Hadoop dwarfs reflect nearly consistent bottlenecks, indicating the Hadoop stack impacts workload behaviors more than other stacks like Spark and TensorFlow.
For Spark dwarfs, which mainly compute in memory, they suffer from a higher percentage of backend bound than that of Hadoop counterparts. Spark Grep, Sample and Sort suffer from more frontend bound and their percentages of backend bound are smaller than the others.
The AI data dwarfs face different bottlenecks both on TensorFlow and Pthreads. Conv and Matmul have the highest IPC (about 2.3) and retiring percentages (about 60\% on TensorFlow). Max pooling, average pooling, and multiplication have extremely low retiring percentages, which has been illustrated in above. However, activation operation like ReLU, sigmoid and tanh suffer from more frontend bound.
%For AI data dwarfs implemented with TensorFlow, convolution (conv) and fully connected (matmul) have about 60\% retiring percentages, they suffer little frontend and bad speculation stalls. For the activation operations like relu, sigmoid and tanh, they reflect similar behaviors which have about 20\% frontend bound percentage.
For AI data dwarfs implemented with Pthread, their main bottleneck is backend bound. They suffer little frontend and bad speculation stalls.

\begin{figure}[!t]
\centering
\includegraphics*[scale=0.85]{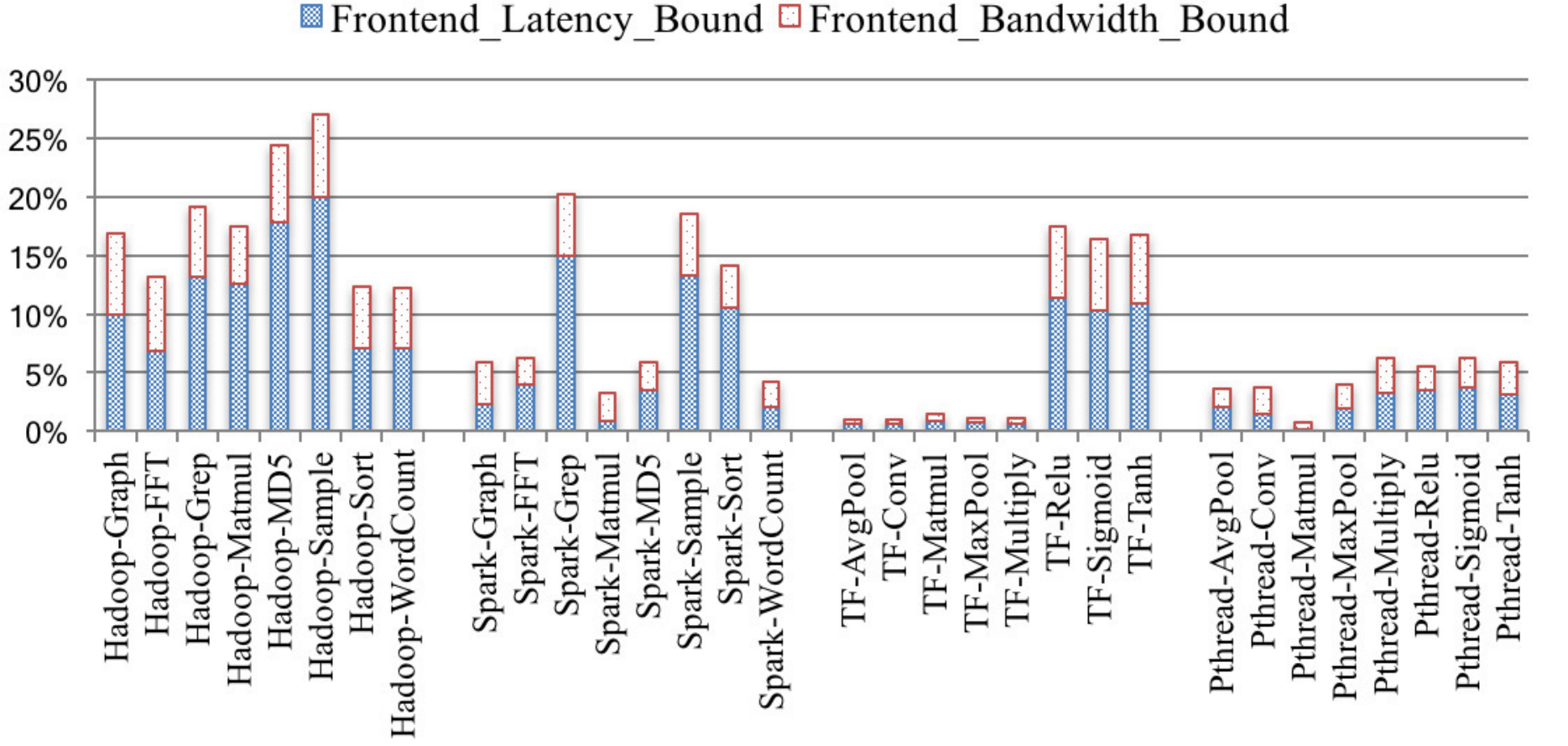}
\caption{The Frontend Breakdown of Data Dwarfs.} %\vspace{1pt}
\label{l2front}
\end{figure}

\begin{figure}[!t]
\centering
\includegraphics*[scale=0.7]{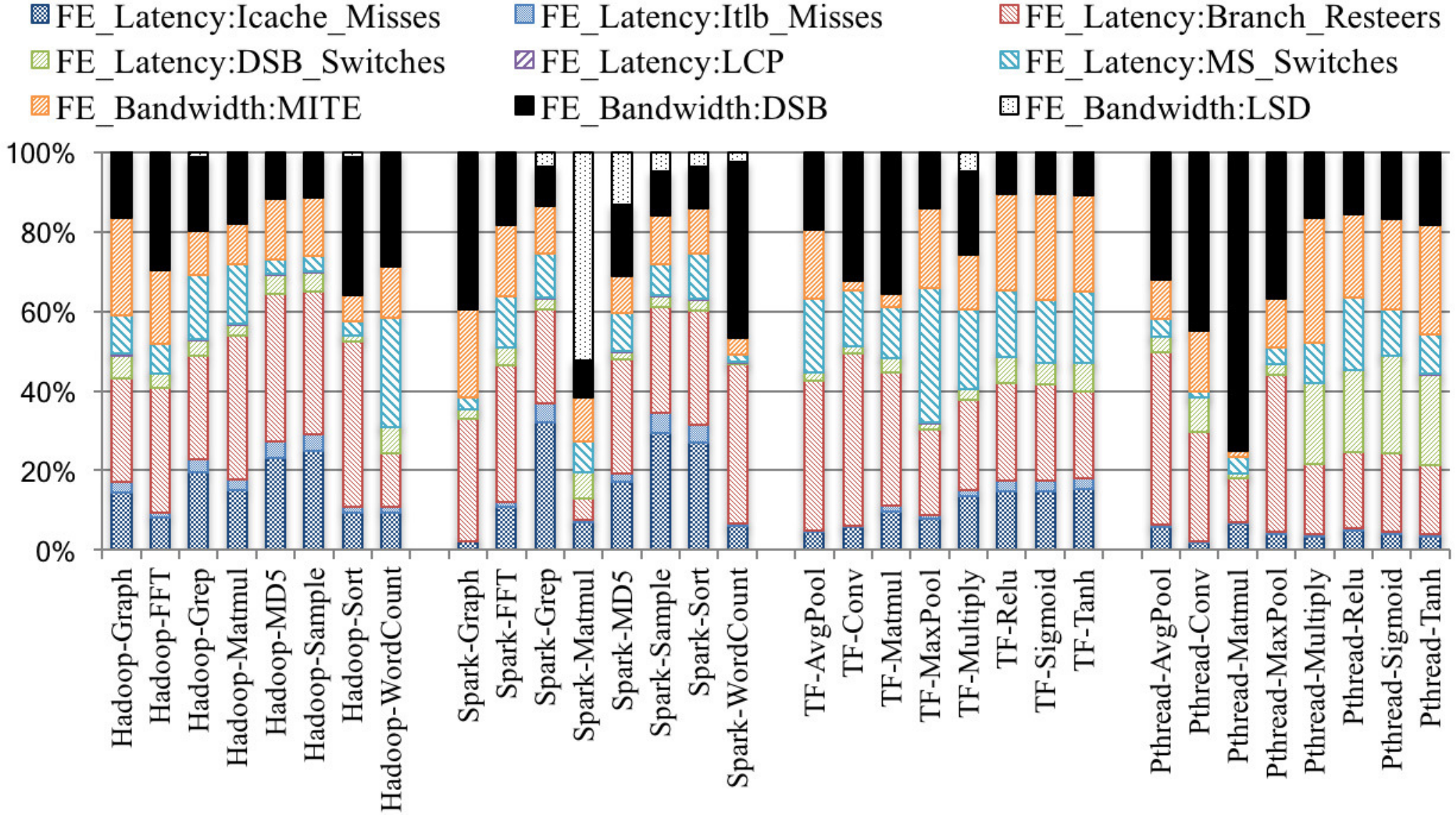}
\caption{The Frontend Latency Breakdown of Data Dwarfs.} %\vspace{1pt}
\label{l3front}
\end{figure}

\textbf{Frontend Bound} Frontend bound can be split into frontend latency bound and frontend bandwidth bound. Among them, latency bound means the frontend delivers no uops to the backend, while bandwidth bound means delivering insufficient uops comparing to the theoretical value. Fig.~\ref{l2front} presents the frontend breakdown of the data dwarfs. We find that the main reason that incurs the frontend stalls is latency bound for almost all dwarfs that suffer severe frontend bound.

We further investigate the reasons for the frontend latency bound and frontend bandwidth bound, respectively. Generally, the frontend latency bound are incurred by six reasons, including icache miss, itlb miss, branch resteers, DSB switches, LCP, and microcode sequencer (MS) switches. Among them, icache miss and itlb miss are instruction cache miss and instruction tlb miss. Branch resteers means the delays to obtain the correct instructions, such as the delays due to branch misprediction. LCP measures the stalls when decoding the instructions with a length changing prefix.
Generally, uops comes from three places, including the decoded uops cache (DSB), legacy decode pipeline (MITE) and microcode sequencer (MS). DSB switches record the stalls caused by switching from the DSB to MITE. MS switches measure the penalty of switching to MS unit.
As for latency bandwidth bound, there are mainly two reasons: the inefficiency of MITE pipeline and the inefficient utilization of DSB cache. Additionally, LSD represents the stalls due to waiting the uops from the loop stream detector~\cite{lsd}.
Fig.~\ref{l3front} lists the latency and bandwidth bound breakdown of all data dwarfs.
%We find that branch resteers is a main reason for almost all data dwarfs except Spark Matmul, whose percentage of frontend bound is only 3\%.
For all data dwarfs except Spark Matmul, we find that branch resteers are the main reason of the high percentage of frontend bound.
Instruction cache miss is more severe on Hadoop and Spark stacks than that on TensorFlow and Pthread stacks, because of the large binary code size. Moreover, MS switch is another significant factor that incurs frontend latency bound. Because big data and AI systems use many CISC instructions that cannot be decoded by default decoder, so they must be decoded by MS unit, and results in performance penalties.
Big data dwarfs implemented with Hadoop and Spark suffer more icache misses than AI data dwarfs.

\begin{figure}[!t]
\centering
\includegraphics[scale=0.6]{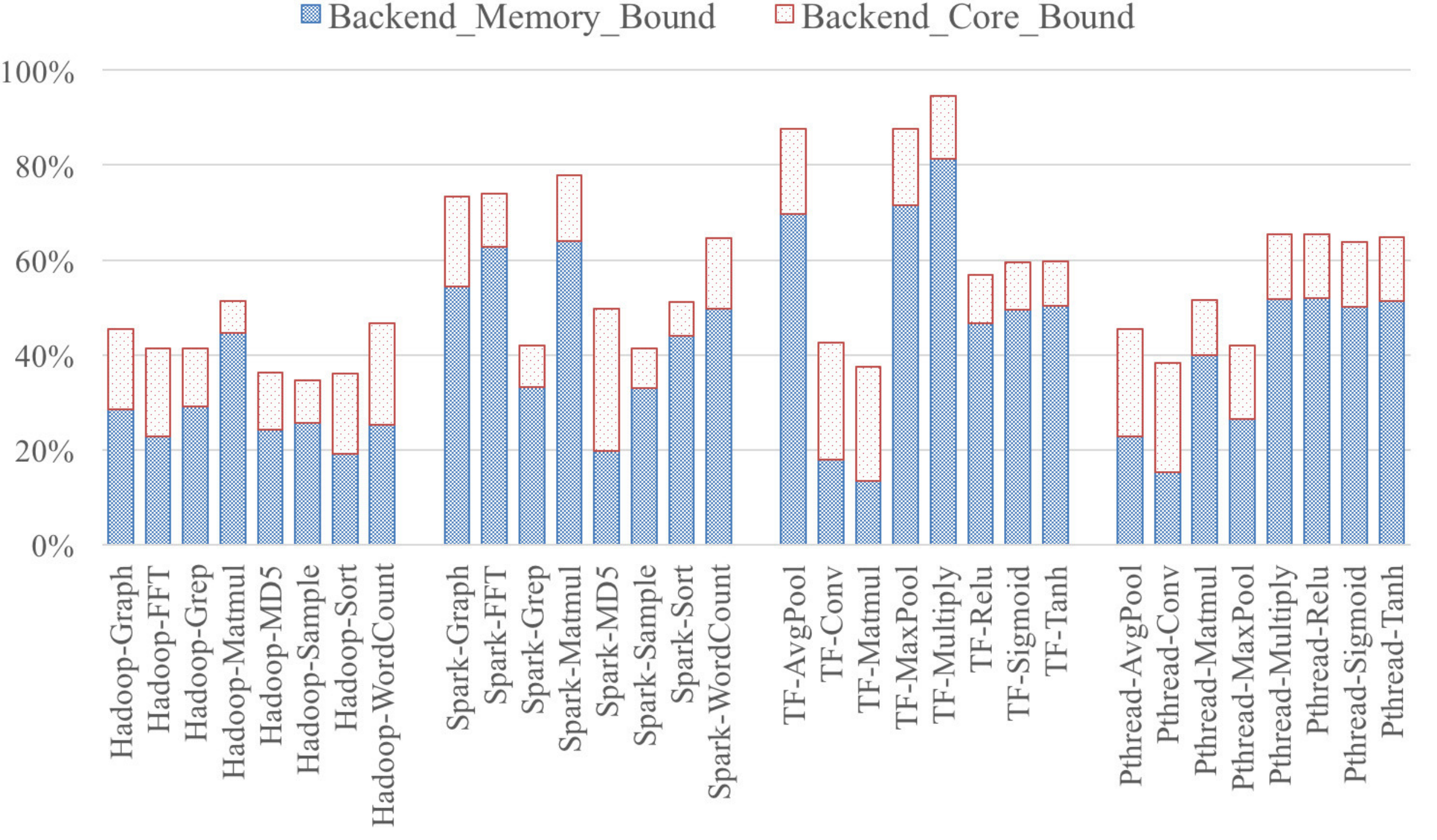}
\caption{The Backend Bound Breakdown of Data Dwarfs.} %\vspace{1pt}
\label{l2back}
\end{figure}

\begin{figure}[!t]
\centering
\includegraphics[scale=0.6]{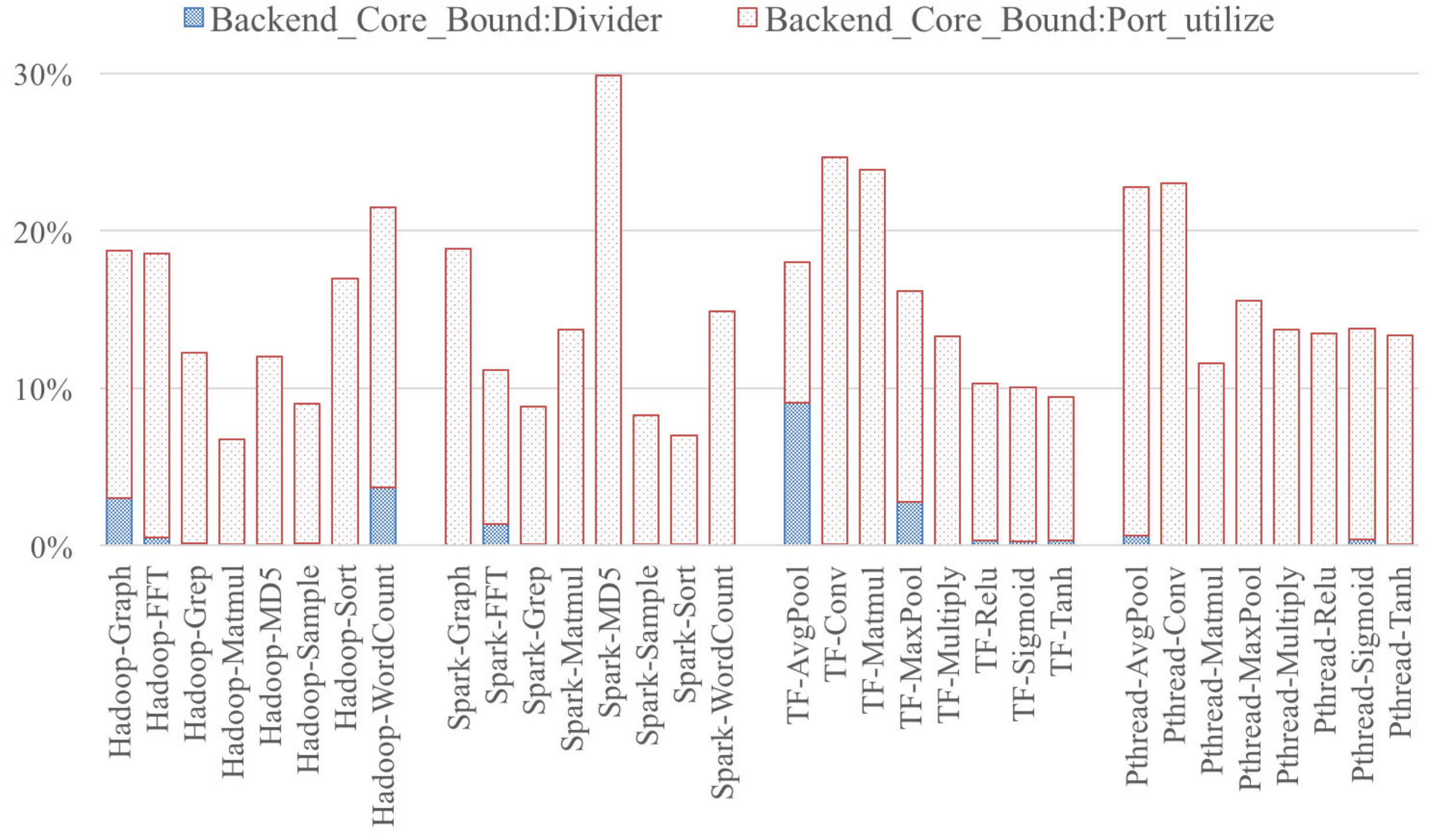}
\caption{The Backend Core Bound Breakdown of Data Dwarfs.} %\vspace{1pt}
\label{l3core}
\end{figure}

\textbf{Backend Bound} Fig~\ref{l2back} presents the backend bound breakdown of data dwarfs, which are split into backend memory bound and backend core bound.
%Backend memory bound represents delays due to memory hierarchy.
Backend memory bound is mainly caused by the data movement delays among different memory hierarchies.
%Backend core bound represents the delays due to a lack of hardware resources (e.g. divider unit) or port under-utilization because of instruction dependencies and execution unit overloading.
Backend core bound is mainly caused by the lackness of hardware resources (e.g. divider unit) or port under-utilization because of instruction dependencies and execution unit overloading.
We find that more than half of these data dwarfs suffer from more backend memory bound than core bound. For each software stack, there is at least one data dwarf that suffer from equal percentages of core bound or even more percentages of core bound than memory bound, such as Hadoop WordCount, Spark MD5, TensorFlow Conv and Pthread Avgpool.
Fig.~\ref{l3core} shows the core bound breakdown. We find that TensorFlow Conv and Hadoop WordCount suffer from significantly long latency of divider unit. While for Spark MD5, which has the highest percentage of backend core bound, mainly suffer from the stalls due to port under-utilization. As for backend memory bound, we find that external memory bound is much severe than level 1, 2, and 3 cache bound for almost all big data and AI dwarfs, indicating that the memory wall~\cite{wulf1995hitting} still exists and need to be optimized.

\begin{figure*}[!t]
\centering
\includegraphics*[scale=0.36]{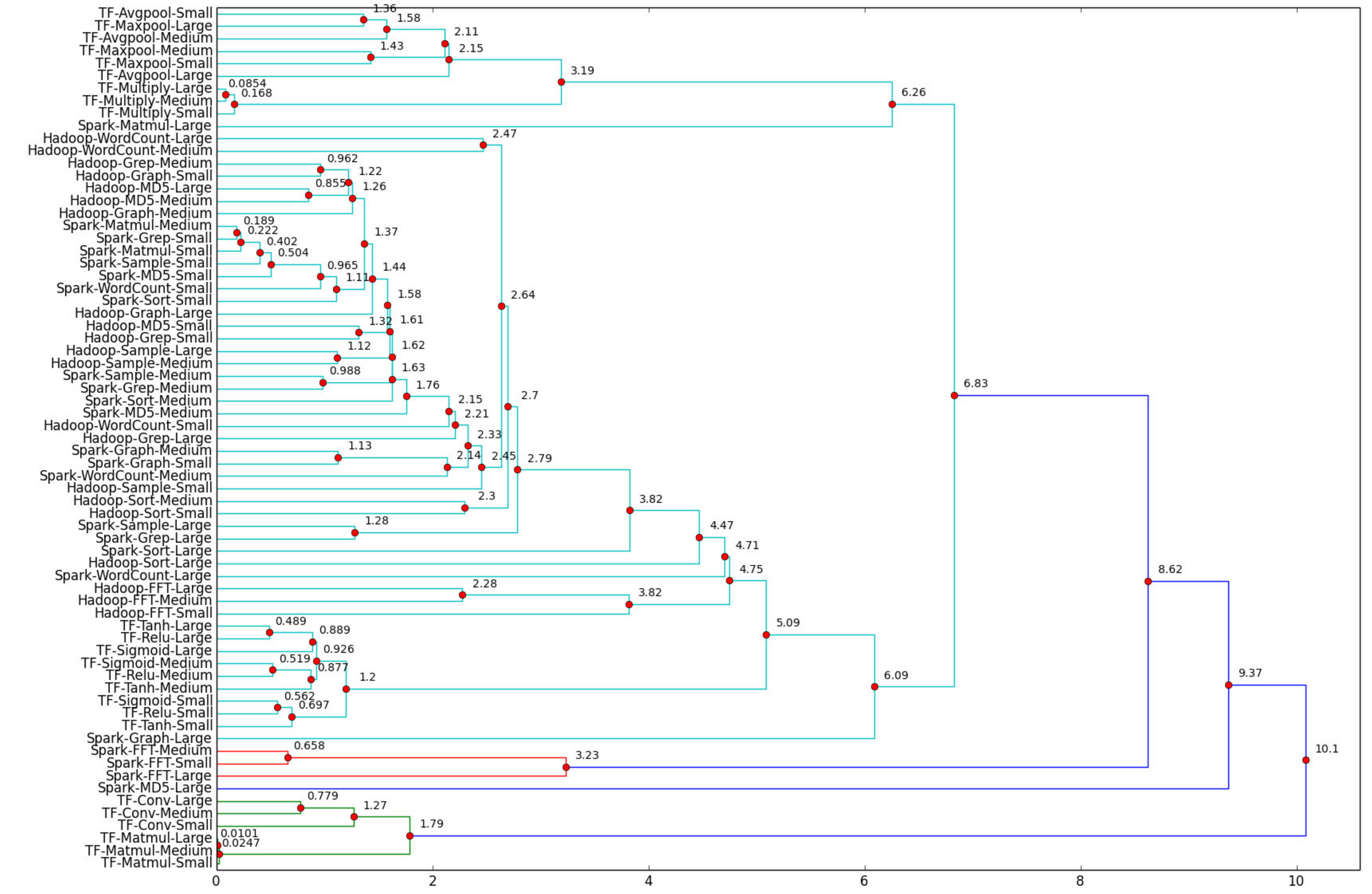}
\caption{Linkage Distance of Data Dwarfs.} %\vspace{1pt}
\label{similar}
\end{figure*}

%\section{Data Impacts on Data Dwarfs}\label{dataimpact}
\section{Impact of Data Input}\label{dataimpact}

In this section, we evaluate the impact of data input on system and micro-architecture behaviors from the perspectives of size, source, type, and pattern. For type and pattern evaluation, we use Sort and FFT as an example, respectively.

\subsection{Impact of Data Size}

%We change the input data size for all the dwarfs we evaluated. For big data dwarfs, each of which has three configurations of \emph{Small}, \emph{Medium}, \emph{Large}.
%For graph dwarf, the \emph{Small}, \emph{Medium}, \emph{Large} sizes are $2^{22}$, $2^{24}$ and $2^{26}$-vertex graph data. For matrix dwarf, we use 100, 1000 and 10000 two-dimensional matrix data with the same distribution and sparsity. For transform dwarf, we use 16384, 32768 and 65536 two-dimension matrix data. For the other big data dwarfs, we use 1, 10 and 100 GB wikipedia text data.
%For AI data dwarfs, we use four configurations with the different input tensor size and channels. They are \emph{(224*224,64), (112*112,128), (56*56,256) and (28*28,512)}. Among them, the first value indicates the dimension of input tensor, the second value indicates the channels. We choose these four configurations because they are widely used in neural network models~\cite{simonyan2014very}. Note that in the rest of paper, we use \emph{dimension-channel} to represent input configurations. For example, we use TF-Conv-224-64 to represent the TensorFlow Conv dwarf with (224*224,64) input tensors.

Based on all sixty metrics spanning system and micro-architecture we evaluated in  Section~\ref{evaluation}, we conduct a coarse-grained similarity analysis using PCA~\cite{jolliffe1986principal} and hierarchical clustering~\cite{johnson1967hierarchical} methods on three data size configurations. Fig.~\ref{similar} presents the linkage distance of all data dwarfs, which indicates the similarity of system and micro-architecture behaviors. Note that the smaller the linkage distance, the more similar the behaviors.
We find that data dwarfs with small data size are more likely to be clustered together. A small data size will not fully utilize the system and hardware resources, hence that they tend to reflect similar behaviors. However, for the dwarf that  is computation intensive and has high computation complexity, even with the large data set, it will be clustered together with small data set. For example, FFTs with three data size configurations are clustered together for both Hadoop and Spark version. AI Dwarfs with TensorFlow implementations are also greatly affected by the input data size. However, they reflect distinct behaviors with big data dwarfs implemented with Hadoop and Spark, with the least linkage distance of 5.09.

\begin{figure}[!t]
\centering
\includegraphics*[scale=0.8]{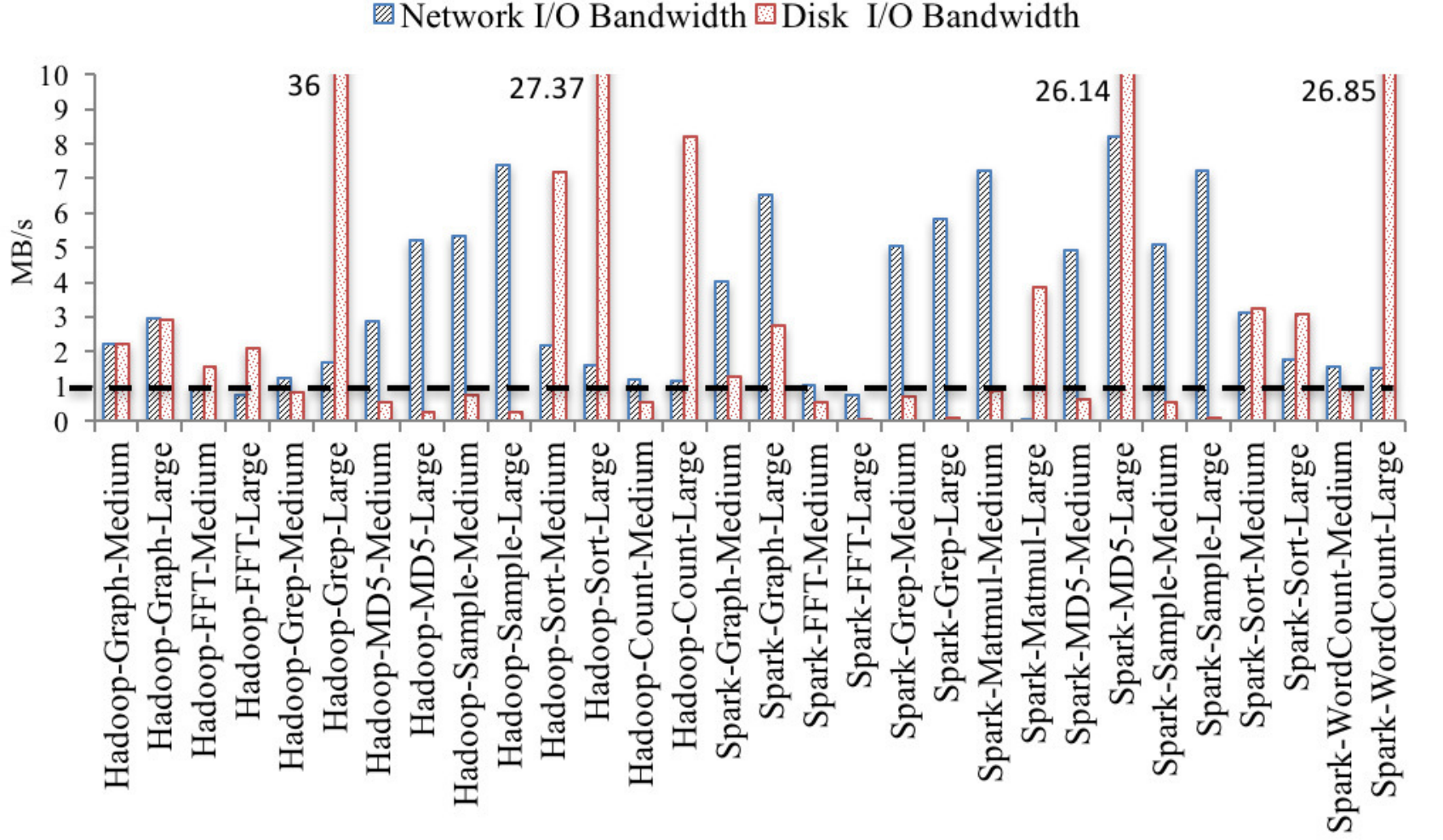}
\caption{Impact of Data Size on I/O Behaviors.} %\vspace{1pt}
\label{ioimpact}
\end{figure}

\textbf{Impact of Data Size on I/O Behaviors}
We evaluate the impact of data size on I/O behaviors using the fully distributed Hadoop and Spark dwarf implementations, as illustrated in Fig.~\ref{ioimpact}. Here we do not report the performance data of the AI dwarfs because the disk I/O behavior is little in neural network modeling,  which we have illustrated in Subsection~\ref{system:exp}. We use the small data input as the baseline, and report the ratio of  the number of the medium or large input divided by that of the small input. The bold black horizontal line in Fig.~\ref{ioimpact} shows the equal line with the small input. That is to say, the value higher than the line means larger I/O bandwidth than the value of the small input.
We find that almost for all data dwarfs, their I/O behaviors are sensitive to the data size. When the data size large enough, the whole data can not be stored in  memory, then the data have to be swapped in and swapped out frequently, and hence put great pressure on disk I/O access. Modern big data and AI systems adopt an distributed manner, with the data storing on an distributed file system, such as HDFS~\cite{shvachko2010hadoop}, the data shuffling or data unbalance will generate a large amount of network I/O.

\begin{figure*}[!t]
\centering
\includegraphics*[scale=0.36]{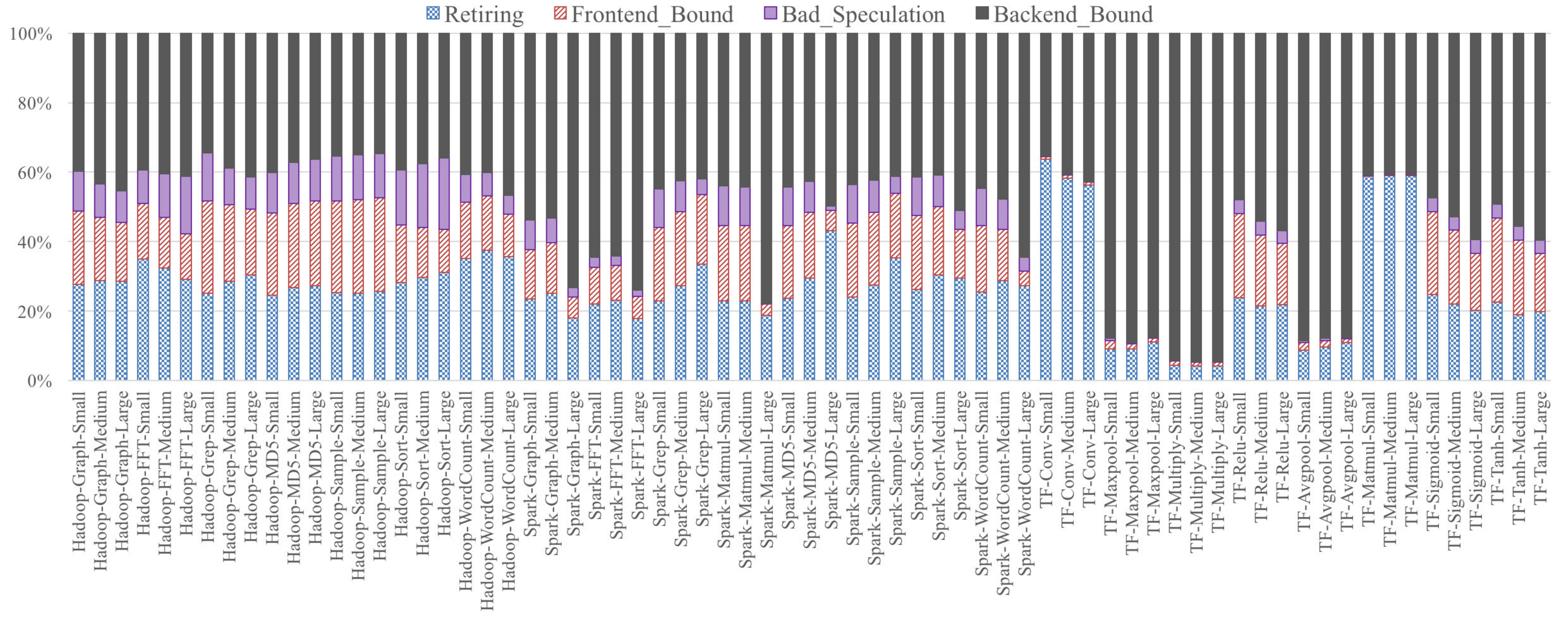}
\caption{Impact of Data Size on Pipeline Efficiency.} %\vspace{1pt}
\label{pipimpact}
\end{figure*}

\textbf{Impact of Data Size on Pipeline Efficiency}
We further measure the impact of data size on pipeline efficiency. As shown in Fig.~\ref{pipimpact}, we find that with the data size increases, the percentage of frontend bound decrease, while the percentage of backend bound increase. For example, Spark Matmul with large input size decrease nearly 20\% of frontend bound and increase more than 30\% of backend bound. As the data size increase, the high-speed cache and even memory are unable to hold all of them, and further incur many data cache misses, resulting in large penalties due to memory hierarchy.

\subsection{Impact of Data Pattern}

\begin{figure*}
\centering
\subfigure[System Behavior with Different Patterns.]{
\label{figP:subfig:a} %% label for first subfigure
\includegraphics[scale=0.45]{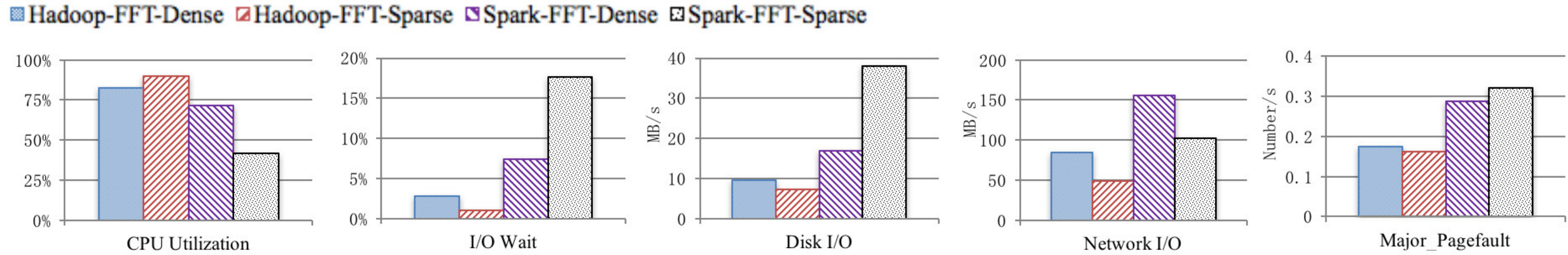}}
\hspace{1in}
\subfigure[Micro-architecture Behavior with Different Patterns.]{
\label{figP:subfig:b} %% label for second subfigure
\includegraphics[scale=0.45]{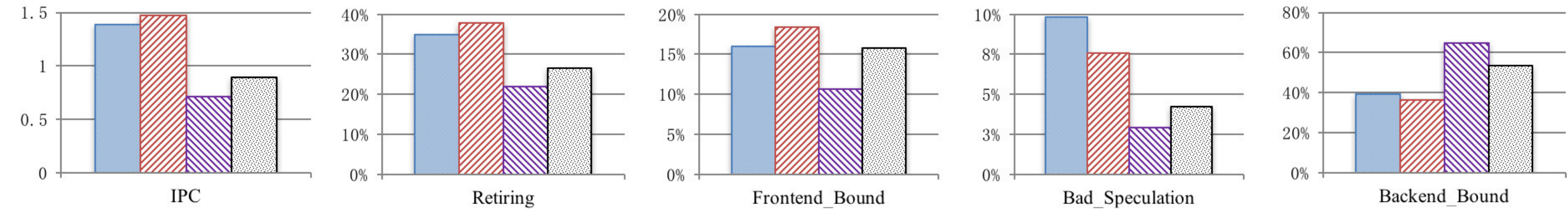}}
\caption{Impact of Data Pattern on Data Dwarfs.}
\label{figP:subfig} %% label for entire figure
\end{figure*}

Data pattern and data distribution impact the workload performance~\cite{xie2018cvr,yilmaz2016autotuning} significantly. To evaluate the impact of data pattern on the dwarfs, we use two different patterns of dense matrix and sparse matrix, to run FFT dwarf as an example. The matrix sparsity indicates the ratio of zero value among all matrix elements. With different sparsity, the data access patterns vary, and thus reflect different behaviors.

We use two 16384$\times$16384 matrixes as the input for the FFT dwarf, with the one having 10\% sparsity and the other one 90\% sparsity.
Fig.~\ref{figP:subfig} shows the impact of data pattern on the data dwarfs from system (Fig.~\ref{figP:subfig:a}) and micro-architecture perspectives(Fig.~\ref{figP:subfig:b}). We find that using the matrix with high sparsity, the network I/O and disk I/O are nearly half of the values using the dense matrix, and the major page fault per second is almost the same. Spark dwarfs suffer from more I/O pressure than Hadoop dwarfs. As for pipeline bottlenecks, sparse data input incurs more frontend stalls while less backend stalls.

\subsection{Impact of Data Type and Source}

\begin{figure*}
\centering
\subfigure[System Behavior with Different Types.]{
\label{fig:subfig:a} %% label for first subfigure
\includegraphics[scale=0.45]{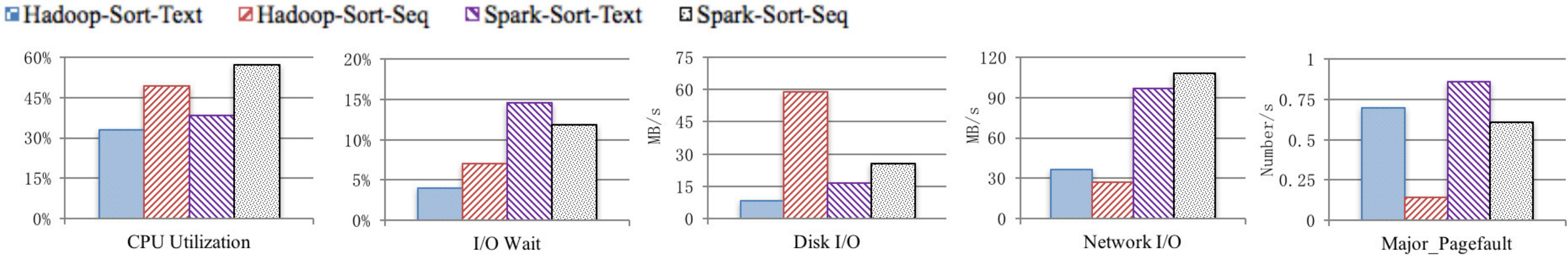}}
%\hspace{-3in}
\subfigure[Micro-architecture Behavior with Different Types.]{
\label{fig:subfig:b} %% label for second subfigure
\includegraphics[scale=0.45]{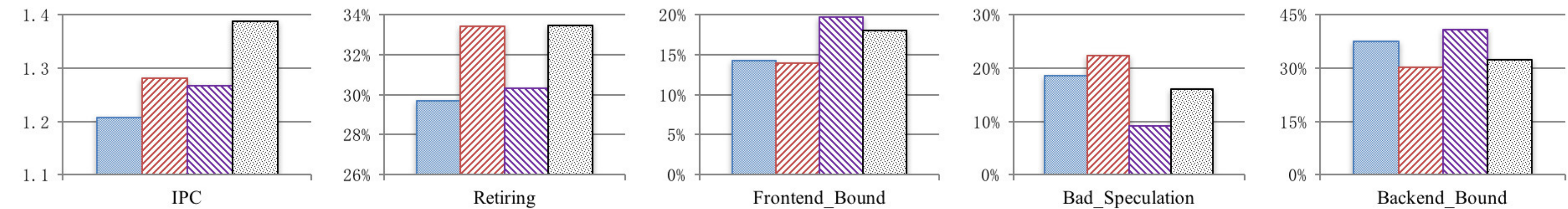}}
\caption{Impact of Data Type on Data Dwarfs.}
\label{fig:subfig} %% label for entire figure
\end{figure*}

Data types and sources are of great significance for read and write efficiency~\cite{eeckhout2003quantifying}, considering their storage format and targeted scenarios, such as the supports for splitable files and compression level.
To evaluate the impact of the data type and source on system and micro-architecture behaviors, we use two different data types for Sort dwarf, with the same data size of 10 GB. two types are un-structured wikipedia text data and semi-structured sequence data. Wikipedia text file is laid out in lines and each line records an article content. Sequence files are flat files that consist of key and value pairs, stored in binary format. Fig.~\ref{fig:subfig} lists the impact of data type on data dwarfs from the system (Fig.~\ref{fig:subfig:a}) and micro-architecture aspects (Fig.~\ref{fig:subfig:b}). We find that the difference between using text type and sequence type ranges from 1.12 times to 7.29 times from the system aspects. 
Using text data type, the CPU utilization is lower than using sequence data, which indicates that using sequence data has better performance.
Both Hadoop Sort and Spark Sort suffer more major page faults and further impact the execution performance, because of page loads from disk. Note that we use the major page fault number per second in Fig.~\ref{fig:subfig} and the total number during the running process is about 100 to 200. 
%However, Sort with sequence data type suffers more interrupts and context switches than with text data type. 
Even with the same amount of data size, their network I/O and disk I/O bandwidth still have a great difference. We find that the sequence format have larger requirements for I/O bandwidth than the text format.
From the micro-architecture aspect (Fig.~\ref{fig:subfig:b}), Sort with different data types reflect different percentages of pipeline bottlenecks. With the text format, backend bound bottleneck is more severe, especially backend memory bound, which indicates that they waste more cycles to wait for the data from cache or memory.

\section{Related Work}

Our big data and AI dwarfs are inspired by previous successful abstractions in other application scenarios.
The \emph{set} concept in relational algebra~\cite{codd1970relational} abstracted five primitive and fundamental operators, setting off a wave of relational database research. The set abstraction is the basis of relational algebra and theoretical foundation of database.
Phil Colella~\cite{colella2004defining} identified seven dwarfs of numerical methods which he thought would be important for the next decade. Based on that, a multidisciplinary group of Berkeley researchers proposed 13 dwarfs which were highly abstractions of parallel computing, capturing the computation and communication patterns of a great mass of applications~\cite{asanovic2006landscape}.
National Research Council proposed seven major tasks in massive data analysis~\cite{council2013frontiers}, which they called giants. These seven giants are macroscopical definition of problems in massive data analysis from the perspective of mathematics, while our eight classes of dwarfs are main time-consuming units of computation in the Big Data and AI workloads.

Application kernels~\cite{bailey1991parallel,dongarra2003linpack} also aim at scaling down the run time of the real applications, while preserving the main characteristics of the workload.
Consisting of the major function of the application, Kernel tries to cover the bottleneck of the real application.
But kernel is still hard to understand the complex and diversity big data and AI workloads~\cite{bailey1991parallel,lilja2005measuring}.
Other than that, kernel mainly focuses on the CPU and memory behaviors, and pays little attention to the I/O, which is also important for many real applications, especially in an era of data explosion.
%Our data dwarf studies the classes of units of computation within real applications, hence that introduces more opportunities for more detailed workload characterization.
%So there is a gap that is not only detail enough to understand the basic operations in a real application systematically, but also taking the I/O in consider.

%There are several successful benchmarks constructed based on the above discussed abstractions.
%TPC-C~\cite{tpcc} proposed the concepts of functions of abstraction and functional workload model, articulated around five kinds of transactions that frequently appeared in OLTP domain~\cite{chen2014tpc}, making it a popular yardstick.
%HPCC~\cite{luszczek2006hpc} is a benchmark suite for high performance computing, which consists of seven basic tests, concentrating on different computation, communication and memory access patterns.
%These successful stories demonstrate the necessity of identifying big data dwarfs and importance of constructing dwarf-based big data (simulation) benchmarks.
%constructing a benchmark suite for simulation based on big data dwarfs.

\section{Conclusions}

In this paper, we answer what are abstractions of time-consuming units of computation
in big data and AI workloads.
We identify eight data dwarfs among a wide variety of big data and AI workloads, as a pipeline of units of computation performed on initial and intermediate data, including Matrix, Sampling, Logic, Transform, Set, Graph, Sort and Statistic computation.
%For each dwarf, we implement the dwarf components  using diverse software stacks.
We implement the data dwarfs for big data and AI separately, including the big data dwarf implementations using Hadoop, Spark, Pthreads, and the AI data dwarf implementations using TensorFlow, Pthreads, considering the impact of data type, data source, data size, and data pattern.
From the system and micro-architecture perspectives, we comprehensively characterize the behaviors of data dwarfs and identify their bottlenecks. Moreover, we measure the impact of data type, data source, data pattern and data size on their behaviors.

%%%%%%%%reference example%%%%%%%%%%%%%%
%\begin{thebibliography}{10}

%\bibliographystyle{ieeetr}
%\bibliography{sample-bibliography}

\end{document}